
\input harvmac.tex

\def\sq{{\vbox {\hrule height 0.6pt\hbox{\vrule width 0.6pt\hskip 3pt
   \vbox{\vskip 6pt}\hskip 3pt \vrule width 0.6pt}\hrule height 0.6pt}}}
\def\half{{1\over2}}

\def\apm{\alpha^{\prime}}
\def\emnpl{\epsilon_{\mu \nu \lambda \rho}}

\def\g{\gamma}

\def\e{\epsilon}

\def\tr{{\rm Tr}}
\def\ie{{\it i.e.}}

\def\cD{{\cal D}}
\def\a{\alpha}

\def\d{\delta}
\def\g{\gamma}

\def\s{\sigma}

\def\e{\epsilon}
\def\p{\partial}

\lref\schout{K. Schoutens, Nucl. Phys. {\bf B295} [FS21] (1988) 634.}

\lref\ps{M. K. Prasad and C. M. Sommerfield, {\it Exact Classical Solutions
for the 't Hooft Monopole and Julia-Zee Dyon,} Phys. Rev. Lett. {\bf 35}
(1975) 760.}

\lref\multimono{E. Corrigan and P. Goddard, {\it An n-Monopole Solution with
(4n-1) Degrees of Freedom,} Commun. Math. Phys. {\bf 80} (1981) 575.}

\lref\AHS{M.F. Atiyah, N. J. Hitchin, and I. M. Singer, {\it Deformations
of Instantons,} Proc. Natl. Acad. Sci. USA {\bf 74} (1977) 2662.}

\lref\dine{X. G. Wen and E. Witten, {\it Electric and Magnetic
Charges in String Theory,} Nucl. Phys. {\bf B261} (1985) 651 \semi
G. Athanasiu, J. Atick, M. Dine and W. Fischler, {\it Remarks on Wilson Lines,
Modular Invariance and Possible String Relics in Calabi-Yau Compactification,}
Phys. Lett. {\bf B214} (1988) 651.}

\lref\ivan{E. Ivanov, S. Krivinos and V. Levant, Nucl. Phys. {\bf B304} (1988)
601 \semi
           Phys. Lett. {\bf B215} (1989) 689.}

\lref\GNO{P. Goddard, J. Nuyts and D. Olive, Nucl. Phys. {\bf B125} (1977) 1.}

\lref\zumino{B. Zumino, Phys. Lett. {\bf B69}, (1977) 369.}

\lref\toptools{E. Witten, {\it Topological Tools in Ten-Dimensional Physics} in
{\it Unified String Theories}, World-Scientific, 1986.}

\lref\colebook{S. Coleman, {\it The Uses of Instantons} in {\it Aspects of
Symmetry},
Cambridge University Press, 1985.}

\lref\raj{ R. Rajaraman, {\it Solitons and Instantons}, North-Holland 1982.}

\lref\ah{N. S. Manton, {\it A Remark on the Scattering of BPS Monopoles}, Phys.
Lett. {\bf 110B} (1982) 54 \semi M. F. Atiyah and N. J. Hitchin, {\it Low
Energy
Scattering of Non-abelian Monopoles}, Phys. Lett. {\bf 107A} (1985) 21. }

\lref\noforce{N. S. Manton, {\it The Force Between 't Hooft-Polyakov
Monopoles,}
Nucl. Phys. {\bf B126} (1977) 525.}

\lref\sbhole{G. Gibbons and K. Maeda, {\it Black Holes and Membranes in
Higher-Dimensional
Theories with Dilaton Fields}, Nucl. Phys. {\bf B298} (1988) 741 \semi
D. Garfinkle, G. Horowitz and A. Strominger, {\it
Charged Black Holes in String Theory}, preprint UCSB-TH-90-66 (1990).}

\lref\gibb{See G. W. Gibbons, {\it Self-gravitating magnetic monopoles, global
monopoles and black holes}, lectures at the 1990 Lisbon Autumn School and
references
therein for a general discussion of monopoles coupled to gravity.}

\lref\manton{N. Manton, Phys. Lett. {\bf B110} (1982) 54.}

\lref\osborn{H. Osborn, {\it Topological Charges for $N=4$ Supersymmetric
Gauge Theories and Monopoles of Spin $1$}, Phys. Lett. {\bf 83B} (1979) 321.}

\lref\giba{G. W. Gibbons, {\it Antigravitating Black Hole Solitons
with Scalar Hair in N=4 Supergravity}, Nucl. Phys. B {\bf 207}, 337
(1982).}

\lref\DAHA{ A. Dabholkar and J. Harvey, {\it Nonrenormalization of
the Superstring Tension},  Phys. Rev. Lett.
{\bf 63} (1989) 719.}

\lref\DGHR{ A. Dabholkar, G. Gibbons, J. Harvey and F. R. Ruiz,
{\it Superstrings and Solitons}, Nucl. Phys. {\bf B340} (1990) 33.}

\lref\hetsol{ A. Strominger, {\it Heterotic Solitons},
Nucl. Phys. {\bf B343} (1990) 167; E: Nucl. Phys. {\bf B353} (1991) 565.}

\lref\HASTR{ J. Harvey and A. Strominger, {\it Octonionic Superstring
Solitons}, Phys. Rev. Lett. {\bf 66} (1991) 549.}

\lref\ADHM{M. F. Atiyah, V. G. Drinfeld, N. J. Hitchin and Y. I. Manin,
{\it Construction of Instantons}, Phys. Lett. {\bf 65A} (1978) 185.}

\lref\hoof{See e.g. R. Rajaraman, {\it Solitons and Instantons}, North-Holland,
1982.}

\lref\RW{ R. Rohm and E. Witten, {\it The Anti-Symmetric Tensor Field in
Superstring Theory}, Ann. Phys. {\bf 170}, 454 (1986). }

\lref\STR{ A. Strominger, {\it Superstrings with Torsion}, Nuclear
Physics {\bf B274}, 253 (1986). }

\lref\sjrey{S. J. Rey, {\it
Confining Phase of Superstrings and Axionic Strings}, Phys. Rev. {\bf D43}
(1991) 439.}

\lref\horrey{G. Horowitz and S. J. Rey, unpublished.}

\lref\SRSTR{ Y. Park, M. Srednicki and A. Strominger, {\it
Wormhole Induced Supersymmetry Breaking in String Theory}, Phys. Lett. B
{\bf 244}, 393 (1990). }

\lref\duffone{M. J. Duff and J. X. Lu,
{\it Elementary Fivebrane Solutions of D=10
Supergravity}, Nucl. Phys. {\bf B354} (1991) 141.}

\lref\dflusix{M.~J.~Duff and J.~X.~Lu, {\it The Self-Dual Type II-B
Superthreebrane}, CTP-TAMU-29/91 (to appear in Phys. Lett. {\bf B}).}

\lref\BEDR{ E. A. Bergshoeff and M. deRoo, {\it The Quartic Effective
Action of the Heterotic String and Supersymmetry}, Nuclear Phys. B
{\bf 328}, 439 (1989).}

\lref\howpaptwo{P. S. Howe and G. Papadopoulos,
{\it Finiteness of (4,q) Models with
Torsion}, Nucl. Phys. {\bf B289} (1987) 264 and {\it Further Remarks
on the Geometry of Two-Dimensional Non-linear $\sigma$ Models},
Class. Quantum Grav. {\bf 5} (1988) 1647. }

\lref\vnew{P. van Nieuwenhuizen and de Wit, {\it Rigidly and
Locally Supersymmetric
Two-Dimensional Nonlinear Sigma Models with Torsion}, Nucl. Phys. {\bf B312}
(1989) 58.}

\lref\BDHM{T. Banks, L. J. Dixon, D. Friedan and E. Martinec, {\it
Phenomenology
and Conformal Field Theory or can String Theory Predict the Weak Mixing Angle},
{\it Nucl. Phys.} {\bf B299} (1988) 613.}

\lref\BD{ T. Banks and L. Dixon, {\it Constraints on String Vacua with
Space-time Supersymmetry}, Nucl. Phys. {\bf B307} (1988) 93. }

\lref\eguchi{See e.g. T. Eguchi, H. Ooguri, A. Taormina and S. K. Yang, { \it
Superconformal Algebras and String Compactification on Manifolds
with $SU(N)$ Holonomy}, Nucl. Phys. {\bf B315} (1987) 193 \semi
N. Seiberg, {\it Observations on the Moduli Space of Superconformal
Field Theories}, Nucl. Phys. {\bf B303} (1988) 286. }

\lref\hetsigref{A review by C. Callan and L. Thorlacius can be found in {\it
Particles, Strings and Supernovae}, Vol. 2, eds. A Jevicki and C.-I. Tan,
World Scientific (1989).}

\lref\cfmpetal{C. Callan, D. Friedan, E. Martinec and M. Perry, {\it
Strings in Background Fields}, Nuc. Phys. {\bf B} (1985) .}

\lref\alfr{L. Alvarez-Gaume and D. Z. Freedman, {\it Geometrical Structure
and Ultraviolet Finiteness in the Supersymmetric Sigma Model},
Comm. Math. Phys. {\bf 80} 443 (1981). }

\lref\bachas{I. Antoniades, C. Bachas, J. Ellis and D. Nanopoulos, {\it
Cosmological String Theories and Discrete Inflation},
Phys. Lett. B{\bf 211} (1988) 393 and Nucl. Phys. B{\bf 328} (1989) 117.}

\lref\hrs{G. Horowitz, A. Strominger and S. J. Rey, unpublished.}

\lref\rusref{V. Dotsenko and V. Fateev,{\it Conformal Algebra and
Multipoint Correlation Functions in Two-Dimensional Statistical Models},
Nucl. Phys. {\bf B240} (1984) 312.}

\lref\ramzi{R. Khuri, {\it Some Instanton Solutions in String Theory},
Princeton preprint PUPT-1219 (1990).}

\lref\sevrin{A. Sevrin, W. Troost and A. van Proeyen, {\it
Superconformal Algebras
in Two Dimensions with N=4}, Phys. Lett. {\bf B208} (1988) 447.}

\lref\gepner{D. Gepner, {\it Spacetime Supersymmetry in
Compactified String Theory and Minimal Models},
Nucl. Phys. B {\bf 296} (1988) 757.}

\lref\ghr{S. J. Gates, C. M. Hull and M. Rocek, {\it Twisted Multiplets
and New Supersymmetric Nonlinear Sigma Models}, Nucl. Phys. B{\bf 248}
(1984) 157.}

\lref\dufftwo{M. J Duff and J. X. Lu, {\it Remarks on String/Fivebrane
Duality}, Nucl. Phys. {\bf B354} (1991) 129.}

\lref\duffthree{M. J. Duff and J. X. Lu, {\it Strings from Fivebranes},
Phys. Rev. Lett. {\bf 66} (1991) 1402.}

\lref\dufffour{M. J. Duff and J. X. Lu, {\it String/Fivebrane Duality, Loop
Expansions and the Cosmological Constant}, CTP-TAMU-108/90.}

\lref\teinep{R. Nepomechie, Phys. Rev. D{\bf 31} (1985) 1921;
C. Teitelboim, Phys. Lett. B{\bf 176} (1986) 69.}

\lref\foursol{T. Banks, M. Dine, H. Dijkstra and W. Fischler, {\it Magnetic
Monopole Solutions of String Theory,} Phys.Lett. {\bf B212} (1988) \semi
B. Greene, A. Shapere, C. Vafa, and S. Yau, {\it Stringy
Cosmic Strings and Noncompact Calabi-Yau Manifolds,} Nucl. Phys. {\bf B337}
(1990) 1 \semi
M. Cvetic, F. Quevedo and S. J. Rey,
{\it Stringy Domain Walls and Target Space Modular Invariance,}
Phys. Rev. Lett {\bf 67} (1991) 1836 \semi
 J. Harvey and J. Liu {\it Magnetic Monopoles in N=4
Supersymmetric Low-Energy Superstring Theory} EFI-90-82, PUPT-1259 (1991). }

\lref\HLP{ J. Hughes, J. Liu and J. Polchinski,  Phys. Lett.
{\bf 180B} (1986), 370. }

\lref\HULL{ C. Hull and E. Witten,  Phys. Lett. B {\bf 160}
(1985) 398 }

\lref\MOOL{ C. Montonen and D. Olive, {\it Magnetic
Monopoles as Gauge Particles?}, Phys. Lett. {\bf 72B} (1977) 117. }

\lref\hupo {J. Hughes and J. Polchinski, Nucl. Phys. B{\bf278}, 147 (1986).}

\lref\TOW{ P. K. Towsend {\it Supersymmetric Extended Solitons},
 Phys. Lett. {\bf 202B} (1988) 53. }

\lref\WO {E. Witten and D. Olive, {\it Supersymmetry Algebras that
include Topological Charges}, Phys. Lett. {\bf 78B} (1978) 97. }


\lref\GiStr{S.B. Giddings and A. Strominger, {\it Exact black fivebranes in
critical superstring theory,} Phys. Rev. Lett. {\bf 67} (1991) 2930.}

\lref\coleman{S. Coleman, {\it The Quantum Sine-Gordon Equation
as the Massive Thirring Model} Phys. Rev. {\bf D11} (1975) 2088.}

\lref\dewit{E.~A.~Bergshoeff and M.~de~Roo, {\it The Quartic Effective
Action of the Heterotic Superstring and Supersymmetry}, Nuc.~Phys.~
{\bf B328} (1989) 439.}

\lref\cfmpetal{C. Callan, D. Friedan, E. Martinec and M. Perry,
{\it Strings in Background Fields}, Nuc.~Phys. {\bf B262} (1985) 593.}

\lref\asen{A.~Sen, {\it Equations of Motion for Heterotic String Theory
from Conformal Invariance of Sigma Models}, Phys.~Rev.~Lett. {\bf 55}
(1985) 1846.}

\lref\gsanom{M. Green and J. Schwarz, Phys.~Lett. {\bf B151} (1985) 21.}

\lref\monosusy{Something on susy approach to finding monopoles}

\lref\hetsol { A. Strominger, {\it Heterotic Solitons},
Nuc.~Phys.~{\bf B343} (1990) 167.}

\lref\howpaptwo{P. S. Howe and G. Papadopoulos,
{\it Finiteness of (4,q) Models with
Torsion}, Nucl.~Phys.~{\bf B289} (1987) 264; {\it Further Remarks
on the Geometry of Two-Dimensional Non-linear $\sigma$ Models},
Class.~Quant.~Grav.~{\bf 5} (1988) 1647. }

\lref\BDHM {T. Banks, L. J. Dixon, D. Friedan and E. Martinec,
{\it Phenomenology and Conformal Field Theory or, Can String Theory
Predict the Weak Mixing Angle?}, Nucl.~Phys.~{\bf B299} (1988) 613.}

\lref\BD{ T. Banks and L. Dixon, {\it Constraints on String Vacua with
Space-time Supersymmetry}, Nucl. Phys. {\bf B307} (1988) 93. }

\lref\alfr{L. Alvarez-Gaume and D. Z. Freedman, {\it Geometrical Structure
and Ultraviolet Finiteness in the Supersymmetric Sigma Model},
Comm. Math. Phys. {\bf 80}
443 (1981). }

\lref\wzwpap{E. Witten, Comm.~Math.~Phys. {\bf 92} (1984) 455. }

\lref\GHS{D. Garfinkle, G. Horowitz, and A. Strominger, Charged black holes
in string theory,'' Phys. Rev. D43 (91) 3140.}

\lref\wsheet{C. Callan, J. Harvey and A. Strominger, {\it Worldsheet
approach to
heterotic solitons and instantons,} Nucl. Phys. {\bf B359} (1991) 611.}

\lref\bhole{G. Horowitz and A. Strominger, {\it Black strings and
$p$-branes,} Nucl. Phys. {\bf B360} (91) 197.}

\lref\withole{E. Witten, {\it On string theory and black holes,} Phys. Rev.
{\bf D44} (1991) 314.}

\lref\wbran{C. Callan, J. Harvey and A. Strominger, {\it
Worldbrane Actions for String Solitons}, preprint PUPT-1244, EFI-91-12 (1991).}

\lref\candelas{P. Candelas, {\it Lectures on Complex Manifolds} in
{\it Superstrings
and Grand Unification, Proceedings of the Winter School on High Energy Physics,
Puri, India, 1988}, T. Pradhan ed., World Scientific Publishers.}

\lref\ryanetal{R. W. Allen, I. Jack, D. R. T. Jones, {\it Chiral $\sigma$
Models
and the Dilaton $\beta$ Function}, Z. Phys. C {\bf 41} (1988) 323 and
references therein.}

\lref\swieca{Proceedings of the Sixth J.~A.~Swieca School of Theoretical
Physics, O.~Eboli ed.}

\Title{\vbox{\baselineskip12pt \hbox{EFI-91-66}}}
{Supersymmetric String Solitons }
\baselineskip=12pt
\bigskip
\centerline{Curtis G. Callan, Jr.}
\centerline{\it Department of Physics, Princeton University}
\centerline{\it Princeton, NJ 08544}
\centerline{\it Internet: cgc@pupphy.princeton.edu}
\medskip
\centerline{Jeffrey A. Harvey}
\centerline{\it Enrico Fermi Institute, University of Chicago}
\centerline{\it 5640 Ellis Avenue, Chicago, IL 60637 }
\centerline{\it Internet: harvey@curie.uchicago.edu}
\medskip
\centerline{Andrew Strominger}
\centerline{\it Department of Physics, University of California}
\centerline{\it Santa Barbara, CA 93106}
\centerline {\it Bitnet: andy@voodoo}

\bigskip
\centerline{\bf Abstract}
These notes are based on lectures given by C. Callan and
J. Harvey at the 1991 Trieste Spring School
on String Theory and Quantum Gravity. The subject is the construction
of supersymmetric soliton solutions to superstring theory. A brief
review of solitons and instantons in supersymmetric theories is presented.
Yang-Mills instantons are then used to construct soliton solutions to heterotic
string theory of various types. The structure of these solutions is discussed
using low-energy field theory, sigma-model arguments, and in one case
an exact construction of the underlying superconformal field theory.

\Date{November 1991}
\eject

\newsec{Introduction}

The theme of most of the lectures given at this school
has been the study of simplified or ``toy'' models of string theory.
The motivation of course is to better understand what a fundamental and
non-perturbative formulation of string theory should look like through
the use of simple models.
These lectures describe a different, but hopefully complementary, approach
to the study of non-perturbative string theory through the development of
semi-classical techniques.

There are several reasons why this is an interesting
enterprise.  First, in realistic unified string theories there is
little hope that we will obtain exact non-perturbative results.  As in most
field theories, probably the best we can hope for is to obtain approximate
non-perturbative results through the use of semi-classical techniques.
Still, even achieving this limited goal is bound to teach us many
interesting things about the structure of string theory, as it has in field
theory.
Second, there are a number of fascinating technical issues in conformal
field theory which must be resolved before this program can be carried out.
These include the proper treatment of collective coordinates and the
proper definition of the mass or action in string theory. Related issues
may also arise in the treatment of back reaction in the exact black hole
solution \withole\ discussed in the lectures of H. Verlinde at this
school. Third, the solitons are in many cases the endpoints of black
hole (or black $p$-brane) evaporation, and
they are therefore an essential ingredient in resolving the fascinating
issues
surrounding Hawking radiation and coherence loss in string theory.
Finally, we may learn something dramatically new about string theory through
the study of semi-classical solutions. For example, the conjecture \hetsol\
that strongly coupled string theory is dual to weakly coupled fivebranes
would have fascinating consequences if true.

The study of semi-classical string theory is still in its infancy. We do not
yet
know how to classify semi-classical solutions directly through
stringy topological invariants nor do we have all the techniques necessary
for extracting physical quantities by expanding string theory about
these solutions. What has been accomplished recently is to find solutions
to string theory, some of which can be studied directly as (super) conformal
field theories, which at large distances are solutions to the low-energy
effective string field theory equations of motion and which have the properties
of solitons and instantons. One goal of these lectures is to discuss these
solutions and to indicate how they might be used to study some of the
issues raised above.

In order to make these lectures reasonably self-contained we will start
in Section 2 by
reviewing certain features of solitons in field theory that will be important
in
what follows. Section 3 reviews aspects of Yang-Mills instantons in
supersymmetric theories. The Montonen-Olive conjecture of a
weak-strong coupling duality in $N=4$ Yang-Mills is described in
Section 4, along with comments on its possible relevance to string theory.
Section 5 discusses solitons and instantons in low-energy heterotic string
theory and the special role played by fivebrane solitons. The
geometry and charges of these solutions is explained in Section 6.
Symmetries, non-renormalization theorems,
and other aspects of the underlying worldsheet sigma model are discussed in
Section 7. An algebraic construction of certain symmetric solutions is
given in Section 8, and the appearance of an exotic $N=4$ algebra is
described. Section 9 briefly reviews some additional supersymmetric
solitons. Concluding comments are made in Section 10.

\newsec{Soliton Review}

Roughly speaking, solitons are static solutions of classical field equations
in $D$ space-time dimensions which are localized in $(D-1)-d$ spatial
coordinates and independent of the $d$ other spatial coordinates. The usual
case is $d=0$ and the soliton then has many characteristics of a point
particle.
For arbitrary $d$ the solution is called a $d$-brane and describes an
extended object. In four spacetime dimensions $d=1$ corresponds to a string
and $d=2$ to a membrane or domain wall. Such objects are of interest
in various cosmological scenarios.

Solitons are usually characterized by the following properties \hoof. First,
they are non-perturbative. They are solutions to non-linear
field equations which cannot be found by perturbation of the linearized
field equations. In addition, their mass (or mass per unit $d$-volume)
is inversely proportional to some power of a dimensionless coupling
constant.  As a result, they become arbitrarily massive compared
to the perturbative spectrum at weak coupling. Thus quantum effects
due to exchange of solitons will be non-perturbative effects which vanish
to all orders in perturbation theory. Second, solitons are characterized
by a topological rather than a Noether charge. Finally, soliton solutions
typically depend on a finite number of parameters called moduli which act as
coordinates on the moduli space of soliton solutions of fixed
topological charge.

A simple and standard example which illustrates these features is the
``kink'' solution in $D=1+1$ spacetime dimensions. Start with the
Lagrangian
\eqn\lag{ {\cal L} = - \half \partial_\mu  \phi \partial^\mu \phi - U(\phi) }
with the potential given by $U(\phi)=\lambda (\phi^2 - m^2/\lambda)^2/4$.
The dimensionless coupling in this example is $g \equiv \lambda/m^2$.
The conserved topological charge is
\eqn\topch{Q = \int_{-\infty}^{+\infty} dx j_0}
where $j_\mu = (\sqrt{g}/2) \epsilon_{\mu \nu} \partial^\nu \phi$ is the
conserved
topological current. Thus
\eqn\phidiff{Q = { \sqrt{g} \over 2} (\phi(+\infty) - \phi(-\infty))}
and is $\pm 1$ for a kink (anti-kink) in which $\phi$ varies from the minimum
of
$U$ at $\phi = \mp 1/\sqrt{g}$ at $x=+\infty$ to the minimum at
$\phi= \pm 1/\sqrt{g}$ at $x=+\infty$.

To find the explicit form of the solution we consider the classical
equation of motion for a
static configuration
\eqn\class{\phi'' = {\partial U \over \partial \phi } .}
For a solution with $U$ and $\phi'$ vanishing at infinity we can integrate this
once to find
\eqn\intonce{ \half (\phi' )^2 = U(\phi).}
Integrating this equation with the previous choice of $U$ yields the
kink (anti-kink) solution
\eqn\kinksol{ \phi_{K (\bar K)}(x) = \pm {m \over \sqrt{\lambda}}
                                     \tanh [ m(x-x_0)/\sqrt{2}].}
The energy (rest mass) of this configuration is
\eqn\energy{ E = \int dx \half (\phi')^2 + U(\phi) = {2 \sqrt{2} \over 3} {m
\over g}}
so that the kink mass divided by the mass of the elementary scalar is
proportional
to $1/g$ showing the non-perturbative nature of the solution.

Also, in accordance with the earlier discussion, the solution for fixed $Q$ and
mass depends on a parameter $x_0$, the center of mass coordinate of the
soliton.
In this example the existence of $x_0$ follows from translational
invariance of the underlying field theory.
The presence of $x_0$ introduces an important problem into the
quantization of the theory expanded about $\phi_K$. This is due
to the fact that the quadratic fluctuation operator
\eqn\quadfluc{{\cal O}_2 = {\delta^2 S \over \delta \phi^2 } |_{\phi=\phi_K} =
                       \partial^2 +m^2 -3 \lambda \phi_K^2}
has a zero mode $\eta_0$ given by an infinitesimal translation of $\phi_K$,
$\eta_0 = \partial \phi_K/\partial x_0$. We can think of $\eta_0$ as a tangent
vector to the curve in configuration space given by translation of $\phi_K$.
A zero mode of ${\cal O}_2$ will lead to a divergence in perturbation theory
about $\phi_K$ of the form ${\rm det}^{-1/2} {\cal O}_2 = \infty$.

The solution is to separate out the dependence on $x_0$ by a change of
coordinates in field
space through the introduction of a collective coordinate. Naively we would
expand $\phi$ in the kink sector as
\eqn\phiexp{\phi(x,t) = \phi_K(x-x_0,t) + \sum_{n=0}^\infty c_n(t)
\eta_n(x-x_0)}
where the $\eta_n$ are a complete set of eigenfunctions of ${\cal O}_2$ and the
$c_n(t)$ are time dependent coefficients which act as coordinates in
configuration
space. The collective coordinate method involves a change of coordinates from
the $\{c_n, n=0 \ldots \infty\}$ to $\{x_0(t), c_n(t), n=1 \ldots \infty\}$
which promotes
the modulus $x_0$ to a time-dependent coordinate. The expansion of $\phi$ is
then
given by
\eqn\collexp{\phi(x,t) = \phi_K(x-x_0(t)) + \sum_{n=1}^\infty c_n(t)
\eta_n(x-x_0(t)).}
It is then possible to separate out explicitly the dependence on $x_0(t)$
and to quantize this as the center of mass coordinate of the soliton
and to be left with a well defined perturbation theory for the non-zero modes.

In the more complicated examples we consider later there will
be moduli resulting from symmetries of the underlying theory (translation
invariance,
scale symmetry, gauge symmetry, supersymmetry) as well as moduli which follow
from index
theorems but are not required by the underlying symmetries.

Two trivial modifications of this theory lead to solutions with
somewhat different interpretations.
We can add some spatial dimensions, say two, to obtain a static solution
to $3+1$ dimensional field equations which is independent of two of the
coordinates, $\phi(x_1,x_2,x_3,t) = \phi_K(x_3)$. Such a solution describes
a two-brane or what would be called in cosmology a domain wall.
Alternatively we can remove the time dimension and reinterpret the solution
$\phi_K(x)$ as a solution in Euclidean time of a $0+1$-dimensional theory
(i.e. quantum mechanics) $\phi_K(x) \sim x(t_E)$. The solution is then an
instanton
describing tunneling between two degenerate minima.

A far less trivial modification is to introduce fermions into the theory. Since
we are
interested in solitons in supersymmetric string theories, we introduce a
supersymmetric
coupling to fermions by taking the Lagrangian to be
\eqn\lsusy{ {\cal L} = - \half (\partial_\mu \phi)^2 + \half \bar \psi i
\gamma^\mu
                     \partial_\mu \psi - \half V^2(\phi) -
                     \half V'(\phi) \bar \psi \psi}
where $\psi$ is a Majorana fermion and $V$ is a function chosen so that the
potential
$V^2$ has two degenerate minima. For concreteness we may take
$V=\lambda(\phi^2-a^2)$.
This theory has two chiral supercharges given by
\eqn\superch{Q_\pm = \int dx (\dot \phi \pm \phi')\psi_\pm \mp V(\phi) \psi_\mp
}
where $\psi_\pm$ are the left- and right-handed components of $\psi$.

It was shown in \WO\ that a careful treatment of
boundary terms leads to the supersymmetry algebra
\eqn\topsusy{Q_+^2 = P_+, \quad Q_-^2 = P_-, \quad \{Q_+,Q_-\}= T}
with $P_\pm = P_0 \pm P_1$, and
where the central term $T$ is equivalent to the topological charge defined
earlier, although its precise functional form is different. The existence of
this central term in the supersymmetry algebra in the soliton sector has
important consequences. First, the relation
\eqn\bogrel{P_+ + P_- = (Q_+ + Q_-)^2 -T = (Q_+ - Q_-)^2 +T}
implies a Bogomolnyi bound  , $M \ge T/2$,  where $M$ is the rest mass.
This bound is saturated precisely for those states $|s \rangle$  for which
$(Q_+ \pm Q_-) |s \rangle= 0$, i.e. for states annihilated by some combination
of
the supersymmetry charges. Classically one can check that this does hold for
the kink
solution using \superch\ and \intonce. In fact, \intonce, or rather its square
root could have been derived by demanding that the kink solution preserve
a linear combination of supersymmetries. The fact that demanding unbroken
supersymmetry leads to a sort of ``square root'' of the equations of motion
will be put to great use later.
Also, the fact that the kink solution is annihilated by one
combination of the supercharges implies
saturation of the Bogomolnyi bound. The other combination of supercharges does
not annihilate the kink state but instead produces a fermion zero mode in the
kink background. This is because a supersymmetry variation of a solution to the
bosonic equations of motion if it is nonzero produces a solution to the
fermionic
equations of motion in the bosonic background.

This simple example provides a paradigm for more complicated examples. In most
known examples of solitons in supersymmetric theories one finds that half of
the supercharges annihilate the classical solution leading to saturation of a
Bogomolnyi bound, while the other half acting on the soliton produce fermion
zero modes
in the soliton background\foot{Although this is not always true. A
fascinating counterexample \HASTR\ will be discussed in Section 9.}.
In addition, searching for configurations which preserve
some of the supercharges provides a shortcut to solving the full equations
of motion since the resulting equations are typically first order
as compared to the second order equations of motion.

In the kink example there is a single fermion zero mode.
It also has interesting consequences
for the spectrum of the theory. The mode expansion of the fermion field in the
soliton
sector will have the form
\eqn\fmode{\psi = b_0 f_0 + \sum_{n \ne 0} b_n f_n}
where $f_0$ is the zero mode. Canonical quantization implies that $b_0$ obeys
$\{b_0,b_0\}=1$ which in turn implies that the vacuum must consist of two
degenerate states, $|+\rangle$ and $|-\rangle$ with $|\pm \rangle = b_0 |\mp
\rangle$.
We can think of the bosonic and fermionic zero modes $(x_0,b_0)$ as providing
coordinates on the supermoduli space of zero energy perturbations of the
soliton.

\newsec{Yang-Mills Instantons}
After this brief review of solitons we turn our attention to instantons in
Yang-Mills
theory. These will provide the starting point for the construction of soliton
solutions
to string theory in ten dimensions, so we will quickly review some of the
features
of instanton solutions which are important in the later constructions.

We start with the Euclidean Yang-Mills action
\eqn\eucact{ S_E = {1 \over 2 g^2} \int d^4x {\rm Tr} F_{\mu \nu} F^{\mu \nu}}
and look for solutions of the equations of motion which approach pure gauge at
infinity,
\eqn\pureg{ A_\mu \rightarrow g^{-1} \partial_\mu g .}
Such configurations are labelled by the winding number of the map from $S^3$
into the gauge group $G$ provided by the gauge function $g$ of \pureg\
evaluated
at infinity. For simplicity we take $G=SU(2)$. It is simple to show that
the action obeys a Bogomolnyi bound in this theory, $S_E \ge 8 \pi^2 k /g^2$
where
\eqn\ctwo{k = {1 \over 16 \pi^2 } \int d^4x {\rm Tr} F_{\mu \nu} \tilde F^{\mu
\nu} }
is the winding number and $\tilde F^{\mu \nu} = \half \epsilon^{\mu \nu \lambda
\rho}
                                               F_{\lambda \rho} $. The bound
is saturated for (anti) self-dual gauge fields obeying
\eqn\selfdual{F_{\mu \nu} = \pm \tilde F_{\mu \nu}.}

The interpretation of these solutions is that they correspond to tunneling
between degenerate minima labelled by elements of $\pi_3(SU(2)$.
Applications of Yang-Mills instantons are described in \colebook\ and \hoof.

To study the bosonic zero modes (moduli) of a solution to the self-dual
equations
we consider a fixed solution $A_\mu ^0$ of \selfdual\ and look for
perturbations
$\delta A_\mu$ which preserve \selfdual with a plus sign. Writing $A_\mu =
A_\mu^0 + \delta A_\mu$
gives $F_{\mu \nu} =F_{\mu \nu}^0 + D_\mu \delta A_\nu - D_\nu \delta A_\mu $
where
the covariant derivative $D$ is evaluated using $A^0$.
Equation \selfdual then requires that
\eqn\sdpert{D_\mu \delta A_\nu - D_\nu \delta A_\mu -
                   \emnpl D^\lambda \delta A^\rho =0. }
We are interested in solutions of \sdpert which are not of the form
$\delta A_\mu = D_\mu \Lambda$, that is not pure gauge. The number of such
solutions
can be determined through the use of the Atiyah-Singer index theorem as
discussed in \AHS. Briefly, one considers the following
elliptic complex
\eqn\ahs{0 \mathrel{\mathop{\longrightarrow}^{D^0}} V^0
            \mathrel{\mathop{\longrightarrow}^{D^1}}
            V^1 \mathrel{\mathop{\longrightarrow}^{D^2_+}}
            V^2_+ \mathrel{\mathop{\longrightarrow}^{D^3}} 0 }
where $V^0$ is the space of zero-forms (scalars), $V^1$ is the space of
one-forms
(deformations of the instanton), and $V^2_+$ is the space of self-dual
two-forms.
$D^0$ and $D^3$ are the identity map and the projection map onto anti-self dual
forms
respectively. $D^1=d+A^0$ and $D^2_+$ is the projection onto the self-dual
part of the covariant exterior
derivative (as in \sdpert). It is easy to check that $D^2_+ D^1 = 0$ iff $F =
\tilde F$.
We then define spaces
\eqn\instcoho{H^i = { {\rm Ker} D^{i+1} \over {\rm Im} D^i }}
and set $h^i = {\rm dim} H^i$. $H^1$ is the moduli space we are after, that is
the space of solutions of \sdpert\ modulo solutions which are pure gauge.
The Atiyah-Singer index theorem then gives the alternating sum of the $h^i$
in terms of topological invariants. For $G=SU(2)$ one has
$h^0 - h^1 + h^2 = -8k+3$ for base manifold $S^4$ while for $R^4$ this sum is
simply $8k$. For $G=SU(2)$ in these cases it is easy to show that $h^0=0$,
that is there are no non-trivial solutions of $D_\mu \Phi =0$ (For other groups
and/or manifolds this is not necessarily true. Gauge connections for which $h^0
\ne 0$
are called reducible connections.) It is also possible to
show using a simple positivity
argument that $h^2=0$ so that on $R^4$ the dimension of the moduli space is
just $8k$. In the later constructions it will be important that we are working
on $R^4$ and not $S^4$ and that the total number of moduli is a multiple of
four.

The construction of an Ansatz which contains all $8k$ parameters is
a difficult and subtle mathematical problem whose solution can be found
in \ADHM. For our purposes this will not be necessary.
There is however a construction which yields $5k$ of the parameters and which
plays an important role in the later constructions.

This ansatz involves writing the gauge field in terms of the
derivative of a scalar field $f$ as
\eqn\thooftans{A_\mu =  {\bar \Sigma}_\mu^\nu \partial_\nu {\rm ln} f}
where the $SU(2)$ indices have been suppressed and with $\bar \Sigma_{\mu \nu}$
a two by two anti-symmetric matrix which is anti-self-dual on the
${\mu \nu}$ indices. A simple construction of $\bar \Sigma$ which will be used
later
arises from embedding  the $SU(2)$  gauge group in $SO(4)$. Letting
$m,n=1 \cdots 4$ be $SO(4)$ indices we have
\eqn\sigmadef{{\bar \Sigma}^{mn}_{\mu \nu } =
         \half (\delta^{mn}_{\mu \nu} - \half \epsilon_{ \mu \nu}^{mn} ).}
which is anti-self-dual both on the $\mu,\nu$ indices and on $m,n$.
This latter fact shows that the $\bar \Sigma$ lie in a $SU(2)$ subgroup
of $SO(4)$.

Substituting this ansatz into the equation \selfdual\ results in
a simple equation for $f$
\eqn\bphi{ {1 \over f} \sq f = 0.}
If $f$ is non-singular then it is constant and hence $A_\mu = 0$.
On the other hand, $f$ can be singular and still solve \bphi\
everywhere if the singularities in $\sq f$ are cancelled
by the $1/f$ factor. The general solution of this form is
\eqn\phisol{f(x) = 1 + \sum_{i=1}^k {\rho_i^2 \over (x-x_i^0)^2 }}
and depends on $5k$ parameters $(x_{i \mu}^0,\rho_i)$ which give
the location and the scale size of the instanton. The $3k$ parameters
which are missing correspond roughly to the relative $SU(2)$ orientations
of the $k$ instantons plus an overall global $SU(2)$ orientation.

For $k=1$ the solution \phisol\ yields the gauge field
\eqn\singainst{A_\mu =-2  \rho^2 \bar \Sigma_{\mu \nu}
                  {(x-x^0)^\nu \over (x-x_0)^2 ((x-x^0)^2 + \rho^2).}}
The gauge field \singainst\ is singular as $x \rightarrow x^0$ and does not
approach
pure gauge at infinity. However, by making a large gauge transformation
one obtains a gauge field
\eqn\ainst{ A'_\mu = -2 \Sigma_{\mu \nu} {(x-x_0)^\nu \over (x-x_0)^2 + \rho^2
}}
which does approach pure gauge at infinity with winding
number one and is non-singular. Here $\Sigma_{\mu \nu}$ is the self-dual
analog of $\bar \Sigma_{\mu \nu}$.

We also want to consider the generalization of these solutions to super
Yang-Mills theory. In $3+1$ dimensions the super Yang-Mills action in component
form is obtained by adding the fermion action
\eqn\fermiact{{\cal L}_F =  {\rm Tr} ( \bar \chi i \gamma^\mu D_\mu \chi ) }
to the  Minkowski version of \eucact\ where $\chi$ is a Majorana fermion in
the adjoint representation of the gauge group. The Euclidean continuation
of this action is not straightforward because of the absence of Majorana
fermions
in Euclidean space. This can be dealt with in various ways.
Luckily these subtleties will not concern us because we will
eventually embed the four-dimensional Euclidean theory in a ten-dimensional
Minkowski theory which does have Majorana fermions. We can thus proceed as
usual except that we must remember that left and right-handed fermions are no
longer related by complex conjugation.

To determine whether the instanton solution \ainst\ is supersymmetric we need
to know whether the supersymmetric variation of this solution is non-zero.
The supersymmetry variation of a configuration with vanishing fermion fields
is
\eqn\susyvar{ \delta \chi = F_{\mu \nu} \gamma^{\mu \nu} \epsilon}
where $\epsilon$ is the parameter of the supersymmetry variation.
Under $SO(4)$ $\epsilon$ can be decomposed into irreducible representations
$\epsilon = \epsilon_L + \epsilon_R$ with $\epsilon_L \sim (2,1)$ and
$\epsilon_R \sim (1,2)$ under $SO(4) \equiv SU(2)_L \times SU(2)_R$.

For an instanton with $F_{\mu \nu} = \tilde F_{\mu \nu}$ we have
\eqn\chiraldel{\delta \chi = F_{\mu \nu} \gamma^{\mu \nu} (1+
\gamma^5)\epsilon}
which is vanishing for $\epsilon_L$ but not for $\epsilon_R$.
Just as in the kink example, half of the supersymmetries are
broken by the solution while the other half give rise to fermion
zero modes. Closer examination shows that of the four fermion
zero modes predicted by the Atiyah-Singer index theorem for $k=1$,
only two arise from supersymmetry transformations corresponding to
the broken supersymmetries as described above.
However, at the classical level, this theory
is also conformally invariant and this leads to superconformal transformations
of the instanton background which produce the two remaining fermion zero modes.

For general $k$ the index theorem predicts $4k$ zero modes, only $4$
of which can be constructed as above. There is however an important
pairing of fermion zero modes with the bosonic moduli which holds
for arbitrary $k$. Given a variation $\delta A_\mu \in H^1$  we can
construct a fermion zero mode of the form
\eqn\fromb{\chi_0  = \delta A_\mu \gamma^\mu \epsilon_L}
There is a redundancy in this
pairing so that two bosonic zero modes are associated to every fermion
zero mode. For details see \zumino.

\newsec{Magnetic Monopoles and the Montonen-Olive Conjecture}

Our final field theory example before discussing string theory has to do with
magnetic monopole solutions in Yang-Mills theory and in particular with the
properties of monopoles in
$N=4$ super Yang-Mills theory.

The action for a Yang-Mills Higgs theory with gauge group $SU(2)$ and a
Higgs field $\Phi$ in the adjoint representation is given by
\eqn\ymh{S = \int -\half \tr F_{\mu \nu}F^{\mu \nu} - \half  \tr D^\mu \Phi
D_\mu \Phi
                  - V(\Phi).}
We assume that $V$ has a minimum in which $\langle \tr\Phi^2 \rangle = v^2$
which
breaks $SU(2)$ down to $U(1)$.
For a static configuration the energy is given by
\eqn\energy{ E = \int d^3 x \tr(B^i \pm D^i \Phi)^2
                 + \int d^3 x V(\Phi) + 4 \pi v |Q_M|}
where $v$ is the asymptotic vacuum expectation value of $\Phi$.
The energy thus satisfies a Bogomolnyi bound
\eqn\bogmon{E \ge  4 \pi v |Q_M|}
where
\eqn\qmdef{Q_M = \int \vec B \cdot d \vec S }
is the magnetic charge and $\vec B$ above
is the asymptotic value of the gauge invariant $U(1)$ field strength.
The magnetic charge $Q_M$ is also related through the equations of motion
to the element of $\pi_2(SU(2)/U(1))=Z$ which labels the topological charge
carried by the Higgs field configuration.
The Bogomolnyi bound is saturated iff $V(\phi) \equiv 0$
and $B^i = \pm D^i \Phi$. If we work in this limit of vanishing potential
then we are free to impose as a boundary condition that the vacuum expectation
value of $\Phi$ approach an arbitrary constant $v$ at spatial infinity.
Of course quantum mechanically we expect a potential to be generated
by renormalization even if it is absent classically.

Nonetheless let us work in this limit for the time being. Then the first
order equations $B^i = \pm D^i \Phi$ can be integrated to give explicit
monopole solutions. The charge one solution can be found in \ps,
multi-monopole solutions are discussed in \multimono.
In this limit there is a universal formula for the classical mass of the
particles of the theory given by
\eqn\bogmass{M^2 =  (4 \pi)^2 v^2 (Q_E^2 + Q_M^2)}
where $Q_E$ and $Q_M$ are the electric and magnetic charges of the
particle respectively. For monopoles this is just the statement
that the Bogomolnyi bound is saturated. For the massive gauge bosons
it is just the usual relation between the gauge boson mass and the Higgs
vacuum expectation value. The photon is of course neutral and massless
as is the remaining physical Higgs boson due to the vanishing potential.

There are some other curious features of this limit. One is the fact
that the static force between two monopoles of like charge or between
two gauge bosons of like charge vanishes. This is due to a cancellation
between a repulsion due to photon exchange and an attraction due
to massless Higgs boson exchange \noforce.

These facts as well as some others led Montonen and Olive  \MOOL,
following ideas of \GNO\
to conjecture the existence of a ``dual'' formulation of the theory
in which electric and magnetic charges are exchanged and in which
the the magnetic monopoles become the gauge bosons and the gauge bosons
arise as solitons of the dual theory. In two spacetime dimensions
the Thirring model -- sine-Gordon duality provides an example
where topological and Noether charges are exchanged in the theory and
its dual, but this duality relies on the peculiar properties of
two dimensions.

One stumbling block to this more ambitious conjecture in $3+1$
dimensions  is the fact that
monopoles appear to have spin zero while gauge bosons of course
have spin one. In addition, the vanishing of the potential is not natural
from the quantum point of view so there is no reason to expect the mass
formula to be exact when quantum corrections are included.

This second objection can be overcome by embedding the theory in
$N=2$ super Yang-Mills theory. Then the charges $Q_E$ and $Q_M$ appear
as central charges in the supersymmetry algebra as in the simpler
kink example and the mass formula \bogmass\ is exact for supersymetric states
\WO.
However an inspection of the fermion zero modes and the resulting monopole
spectrum in this theory shows that the monopole states fill out a matter
supermultiplet consisting of spin zero and spin one-half states.
In order to construct monopoles with integer spin it is necessary
to extend the supersymmetry to $N=4$, the maximal allowable global
supersymmetry in $3+1$ dimensions. This theory has a number of remarkable
features. First, the structure of the fermion zero modes is such that
the monopole super multiplet now coincides with the gauge supermultiplet
and includes states of spin $1$, $1/2$, and $0$ \osborn.
Second, the scalar potential has exact flat directions due to supersymmetry
and again the mass formula is exact. Finally, this theory is finite
with vanishing beta-function so that a duality which relates $g \rightarrow
1/g$
can make sense quantum mechanically at all scales. Thus in this special
theory all of the simple objections to the existence of the sort of duality
suggested by Montonen and Olive disappear! Of course this is a far cry
from showing that such a duality actually holds, but the evidence is suggestive
enough that the idea is well worth pursuing.

Finally, it is perhaps worth mentioning that the $N=4$ theory and the Montonen-
Olive conjecture may have some ties to ten-dimensional physics and hence to
string
theory. For one thing, the $N=4$ theory can be obtained by dimensional
reduction
of $N=1$ super Yang-Mills in ten dimensions. Second, the Montonen-Olive
conjecture
for more general gauge groups says that the dual gauge group should have
a weight lattice dual to the weight lattice of the original group.
Self-dual lattices of course play a crucial role in ten-dimensional heterotic
string theory with the gauge groups $SO(32)/Z_2$ and $E_8 \times E_8$ with
self-
dual lattices being singled out by anomaly cancellation. Finally, a stringy
analog of the Montonen-Olive conjecture\hetsol\ will be briefly discussed
in the following section.

\newsec{Low-Energy Heterotic String Theory}

Having discussed soliton and instanton solutions of various supersymmetric
field theories we would like to generalize these considerations to string
theory.
Let us first discuss the problem of finding string solitons via the ``strings
in background fields'' spacetime approach. The beta functions for strings
propagating in a background of massless fields are the equations of motion
of a certain master spacetime action which can be computed as an expansion
in the string tension $\apm$. For the heterotic string, the leading terms in
this action are identical to the $D=10$, $N=1$ supergravity and super
Yang-Mills action. The bosonic part of this action reads
\eqn\action{S = {1 \over {\apm}^4}\int d^{10}x \sqrt{-g} e^{-2 \phi} \left(
                R + 4(\nabla \phi)^2 - {1 \over 3} H^2 -
                {\apm \over 30} {\rm Tr} F^2 \right), }
where the three-form antisymmetric tensor field strength is related to the
two-form potential by the familiar anomaly equation \gsanom
\eqn\anom{H=dB +\apm\left(\omega^L_3(\Omega_-)-{1\over 30}
			\omega^{YM}_3(A)\right)+\ldots}
(where $\omega_3$ is the Chern-Simons three-form) so that
\eqn\curlh{dH=\apm (trR \wedge R- {1 \over 30}Tr F \wedge F).}
The trace is conventionally normalized so that
$TrF \wedge F = \sum_i F^i \wedge F^i$ with $i$ an adjoint gauge group index.
An important, and potentially confusing, point is that the connection
$\Omega_\pm$ appearing in \anom\ is a non-Riemannian connection related
to the usual spin connection $\omega$ by
\eqn\connect{ \Omega_{\pm M}^{AB} = \omega_M^{~AB} \pm  H_M^{~AB}. }
Since the antisymmetric tensor field plays a crucial role in all of our
solutions, this subtlety will be crucial.

Rather than directly solve the equations of motion for this action, it is
much more convenient to look for bosonic backgrounds which are annihilated by
{\it some} of the N=1 supersymmetry transformations (only the vacuum is
annihilated by all the the supersymmetries). Both the kink solution
\kinksol\
and the self-dual equation \selfdual\ could have been found in this manner.
The Fermi field
supersymmetry transformation laws which follow from \action\ are
\eqn\susyvar{ \eqalign{\delta \chi & = F_{MN} \gamma^{MN} \epsilon \cr
                       \delta \lambda & = (\gamma^{M} \partial_M \phi -
                       {1 \over 6} H_{MNP}\gamma^{MNP}) \epsilon  \cr
                        \delta \psi_M & = (\partial_M +
                                           {1 \over 4}\Omega_{-M}^{AB}
                                          \gamma_{AB}) \epsilon,   \cr }}
and it is apparent that to find backgrounds for which all of \susyvar\
vanish, it is only necessary to solve first-order equations, rather than
the more complicated second-order equations which follow from varying
the action. We will shortly construct a simple ansatz for the bosonic fields
which does just this.

Although there are a variety of soliton or instanton solutions of \susyvar\
we could consider, it is useful to first study the simplest and most
natural solutions.
In realistic theories we would consider a spacetime of the form
$M \times K$ with $M$ four-dimensional Minkowski or Euclidean space
and $K$ a six-dimensional compact space which solves the equations
of motion following from \action. Soliton and instanton
solutions involving gauge fields
on such spaces have been studied and a general classification has been
discussed in \toptools. However these solutions are either related to
the usual Yang-Mills instantons or involve the structure of $K$ in
various ways. A simpler but less realistic starting point would
be to ask for solutions to string theory in ten-dimensional Minkowski
space (or Euclidean space in the case of instantons). We will discuss
solutions of this form both because they allow us to discuss solitons
without the complication of discussing compactification and because
the solitons we find may have implications for fundamental string theory.

The solutions of \susyvar\ that we will consider are fivebrane solitons.
The motivation for making such an apparently arbitrary choice
is based on an analogy between
these solitons in string theory and Dirac monopoles in electromagnetism.
In $3+1$ dimensions a point particle naturally couples to a one-form potential
$A$ via
\eqn\myone{ \int_L A \equiv \int d \tau {d x^\mu \over d \tau} A_\mu }
where $L$ indicates the world-line of the particle, determined by
$x_\mu(\tau)$ which gives the particle location in spacetime
as a function of a time-like parameter $\tau$.
This coupling results in an ``electric'' source term in the equation
of motion for the field strength $F=dA$
\eqn\mytwo{ d {}^* F = j_E .}
Using Stoke's theorem the electric charge of the point particle is
\eqn\mythree{ Q_E = \int_{S^2_\infty} {}^* F }
where $S_\infty^2$ is a two-sphere at spatial infinity.

Dirac conjectured that there might exist magnetic monopoles which would
obey a dual version of these equations
\eqn\myfour{ dF=0 \quad \Longrightarrow \quad dF= \tilde j_M }
\eqn\myfive{ d {}^* F = j_E \quad \Longrightarrow \quad d {}^* F =0 }
\eqn\mysix{ Q_E = \int_{S^2_\infty} {}^* F \quad \Longrightarrow \quad
   Q_M = \int_{S^2_\infty} F }
and showed that quantum consistency imposes a quantization condition on the
product of the electric and magnetic charges $Q_E Q_M = 2 \pi n$ with n
an integer.

There is a construction analogous to this in string  theory. Since we
now want to couple to the world-sheet of a string rather than the world-line
of a particle we introduce a two-form potential $B$, called a
Kalb-Ramond field,  which couples to the
string world-sheet $S$ via
$ \int_S B $
and has a three-form field strength $H=dB$. This is the same $H$ as in \anom\
except that for the moment we ignore the Chern-Simons terms in Equation \anom.
The analogs of the previous
equations and their dual forms are then (in D spacetime dimensions)
\eqn\myseven{ dH=0 \quad \Longrightarrow \quad dH= j_M     }
\eqn\myeight{ d {}^* H = j_E \quad \Longrightarrow \quad d {}^* H = 0 }
\eqn\mynine{ Q_E = \int_{S_\infty^{D-3}} {}^* H \quad \Longrightarrow Q_M =
         \int_{S_\infty^3} H. }
Because ${}^*H$ is a $D-3$ form,
it must be integrated over a $D-3$ dimensional manifold. This fits perfectly
with our string picture since a  static infinite string in D spacetime
dimensions has a $D-3$ sphere at spatial infinity. However this also means that
the dual object has a charge measured by integrating $H$ over a
three-dimensional
manifold. So, unless $D=6$, the string and its dual object do not have the
same dimension. For example, consider $D=4$. It is well known that the
two form $B$ is equivalent to a massless scalar field in four dimensions
and that this scalar field has couplings analogous to the usual axion field.
The ``electric'' charge, measured by integrating ${}^* H$ over the $S^1$
at infinity around an infinite string, is just the axion charge of the string.
The dual object has a ``magnetic'' charge obtained by integrating $H$
over a three-sphere at infinity. But this means that the dual object must be
localized
in both space and time, i.e. it is an instanton!
Thus we learn the important fact that in four dimensions
strings are dual, in the Dirac sense,
to instantons. If we move up to $D=10$ as required by
superstring theory, then the dual object is a fivebrane, that is an object
with extent in five spatial dimensions which sweeps out a six-dimensional
world-volume as it evolves in time.

This fivebrane
is also intimately related to Yang-Mills instantons in four dimensions.
One way to see this connection is to note that in string theory we no longer
have $H=dB$ but rather
\eqn\myten{H=dB - { \alpha^\prime \over 30}  \omega_3^{YM} }
(the additional
gravitational contributions which will be discussed shortly). We thus have
the equations
\eqn\myeleven{ d H = - {\alpha^\prime \over 30} Tr(F \wedge F) }
\eqn\mytwleve{ d {}^* H = 0 }
in a region free of electric sources. Comparison with the previous equations
clearly shows that the Yang-Mills topological charge density acts as a
magnetic source term for $H$.

This duality argument doesn't guarantee that the dual objects
actually exist, of course, but we shall see that in fact they do.
The existence of objects which are {\it geometrically} dual to strings
is quite interesting and raises the question of whether they could
also be {\it dynamically} dual as in the conjecture of Montonen and
Olive. This possibility was raised in \hetsol, some issues associated
with and evidence for
this conjecture are discussed in\refs{ \wbran\duffone\dufftwo\duffthree}.
Although the idea is
a tantalizing one, proving or disproving it
seems beyond our current ability.

With this motivation let us attempt to construct a fivebrane solution
to \susyvar .  The supersymmetry
variations are determined by a positive chirality Majorana-Weyl $SO(9,1)$
spinor $\e$. Because of the fivebrane structure, it is useful to note that
$\e$ decomposes under $SO(9,1)\supset SO(5,1)\otimes SO(4)$ as
\eqn\decomp{ 16 \to (4_+,2_+)\oplus  (4_-,2_-)}
where the $\pm$ subscripts denote the chirality of the representations.
Denote world indices of the four-dimensional space transverse to the fivebrane
by $\mu, \nu = 6 \ldots 9$ and the corresponding tangent space indices by
$m,n =6 \ldots 9$. We assume that no fields depend on the longitudinal
coordinates (those with indices $M = 0 \ldots 5$) and that the nontrivial
tensor fields in the solution have only transverse indices. Then the
gamma matrix terms in \susyvar\ are sensitive only to the $SO(4)$ part of $\e$
and, in particular, to its $SO(4)$ chirality.

One immediately sees how to  make the gaugino variation vanish (in what follows
we treat $\e$ as an $SO(4)$ spinor and let all indices be four-dimensional):
As a consequence of the four-dimensional gamma-matrix identity
$\g^{mn}\e_\pm= \mp\half\e^{mnrs}\g^{rs}\e_\pm$ one has
$F_{mn}\g^{mn}\e_\pm= \mp\widetilde F_{mn}\g^{mn}\e_\pm$, where the dual field
strength is defined by $\tilde F_{mn}=\half\e_{mnrs}F^{rs}$. Therefore,
$\delta\chi$ vanishes if
	$F_{mn}=\pm\tilde F_{mn}$ and $\e=(4_\pm,2_\pm)$
which is to say that if the gauge field is taken to be an {\it instanton}, then
$\delta\chi$ vanishes for all supersymmetries with positive $SO(4)$ chirality.
This is just the argument of \chiraldel\ except that the supersymmetry
parameter
now transforms under $SO(5,1)$ as well.

To deal with the other supersymmetry variations, we must
adopt an {\it ansatz} \STR\
for the non-trivial behavior of the metric and antisymmetric tensor fields
in the four dimensions transverse to the fivebrane.
For the metric tensor we write
\eqn\metric{g_{\mu\nu}=e^{+2\phi}\delta_{\mu\nu} , \qquad \mu~,\nu=6\ldots 9}
and for the antisymmetric tensor field strength
\eqn\ast{H_{\mu\nu\lambda}=
		- {\e_{\mu\nu\lambda}}^\sigma\partial_\sigma\phi , }
where $\phi$ is to be identified with the dilaton field. With this ansatz
and the rather obvious vierbein choice $e^m_\mu=\delta^m_\mu~e^{+\phi}$,
we can also calculate the generalized spin connection \connect\ which
appears in \susyvar\ and \anom :
\eqn\consatz{{\Omega_\pm}_{\mu mn}=\delta_{m\mu}\p_n\phi
	-\delta_{n\mu}\p_m\phi\mp{\e_{\mu mn}}^\rho\p_\rho\phi~.}
Now consider the $\delta\lambda$ term in \susyvar . Because of the ansatz,
both terms are linear in $\p\phi$. By standard four-dimensional gamma-matrix
algebra, the relative sign of the two terms is proportional to the $SO(4)$
chirality of the spinor $\e$. We have chosen the sign and normalization
of the ansatz for $H$ so that $\delta\lambda$ vanishes for $\e\in (4_+,2_+)$.
Finally, consider the gravitino variation in \susyvar . A crucial fact,
following from \consatz\ , is that while $\Omega_\pm$ would in general be
an $SO(4)$ connection, with the chosen ansatz it is actually pure $SU(2)$.
To be precise,
\eqn\sutwo{{\Omega_\pm}^{mn}_\mu\g_{mn}\e_\eta=
	2 ({\g_\mu}^\rho\p_\rho\phi)(1\mp\eta 1)\e_\eta~,}
(where $\eta = \pm $)
so that $\Omega_\pm$ annihilates the $(4_\pm,2_\pm)$ spinor. Since \susyvar\
involves only $\Omega_-$, it suffices to take $\e$ to be a {\it constant}
$(4_+,2_+)$ spinor to make the gravitino variation vanish. It can
be shown that {\it all}
supersymmetric fivebrane configurations are of the form
given by the ansatz
\metric, \ast.

Putting all this together, we see that if we choose the gauge field to be any
instanton and fix the metric and antisymmetric tensor in terms of the dilaton
according to the above ansatz, then the state is annihilated by all
supersymmetry variations generated by a spacetime constant $(4_+,2_+)$ spinor.
Thus, {\it half} of the supersymmetries are unbroken, and the other half,
by standard reasoning, will be associated with fermionic zero-modes
bound to the soliton.

The one unresolved question concerns the functional form of the dilaton field.
Notice that the ansatz for the antisymmetric tensor was given in terms of
its three-form field strength $H_{\mu\nu\rho}$, rather than its two-form
potential
$B_{\mu\nu}$. This is potentially inconsistent, since the curl of the field
strength must satisfy the anomalous Bianchi identity \curlh. Within the
ansatz \ast, the curl of $H$ is given by
\eqn\anomeq{dH=-{1 \over 2} {}^*\sq e^{-2\phi}= {\apm }
	({\rm tr} R\wedge R-{1\over 30}{\rm Tr} F\wedge F)~.}
where $\sq$ is the flat space laplacian and $*$ is the four-dimensional
Hodge
dual. This equation can be solved perturbatively in $\apm$ beginning
with the instanton solution for $F$.
Equation \anomeq\ then implies that $\p\phi\sim O(\apm)$
which in turn implies that $R\sim O(\apm)$. Therefore, to
leading order in $\apm$, one is entitled to drop the $R\wedge R$ term in
\anomeq. Substituting the explicit gauge field strength for an instanton
of scale size $\rho$, one obtains the following dilaton solution:
\eqn\stromdil{e^{+2\phi}=e^{+2\phi_0}+
                8\apm{(x^2+2\rho^2)\over(x^2+\rho^2)^2}+O({\apm}^2)~.}
The metric and antisymmetric tensor fields are built out of this dilaton field
according to the spacetime-supersymmetric ansatz of \metric\ and \ast .
One can examine higher-order in $\apm$ corrections to the beta
functions and verify that the solution must receive corrections. At the
same time, one can examine subleading corrections to the supersymmetry
transformations \dewit\ and verify that it is possible to maintain
spacetime supersymmetry in the $\apm$-corrected solution. However one would
like to know if there is an $exact$ solution of string theory
which agrees with this one at long distances. In fact the existence of such
an exact solution can be demonstrated using low-energy spacetime
supersymmetry. As these notes focus on the worldsheet point of view, we
refer the reader to the original literature \wbran\ for details.

While \stromdil\ is in some sense the most obvious supersymmetric
fivebrane solution, there is in fact a different solution with
more worldsheet supersymmetry. This extra worldsheet supersymmetry
will be seen to have powerful consequences: the solution is exact
without any $\apm$ corrections. This symmetric solution is characterized,
from the spacetime point of view, by $dH=0$.
This condition requires,
according to \curlh\ , that the curvature $R(\Omega_-)$
should cancel against the instanton Yang-Mills field $F$. We will take
the instanton to be embedded in an $SU(2)$ subgroup of the gauge group
(this is always the lowest-action instanton), so what is needed is that
$\Omega_-$ be a self-dual $SU(2)$ connection. In fact $\Omega_-$ is
manifestly an $SU(2)$ connection,
so the only issue is self-duality. Given the special ansatz
and coordinate system of \metric\ , it is easy to calculate the
curvature of $\Omega_\pm$.
It is then a simple arithmetical exercise to show that, in four dimensions
and under the condition that $\sq e^{2\phi} =0$,
$\Omega$ is self-dual:
\eqn\omslfdl{R(\Omega_\pm)^{mn}_{\mu\nu}=
  \mp\half{\e_{\mu\nu}}^{\lambda\sigma} R(\Omega_\pm)^{mn}_{\lambda\sigma}~.}
Since a self-dual $SU(2)$ connection is an instanton connection, it will
be possible to choose a gauge instanton which exactly matches the
``metric'' instanton $\Omega_-$ and makes the {\it r.h.s} of \curlh\
vanish, thus making the whole solution self-consistent.
%
%
In the next section, we will explore the qualitative properties of the
solutions which we have constructed in the above rather roundabout manner.

A feature of the above development which could cause confusion is the intricate
interplay of the two non-Riemannian connections $\Omega_{(\pm)}$. To refresh
the reader's memory, we will summarize the essentials of this phenomenon
(we denote the $(4_+,2_+)$ spinor by $\e_+$): The gravitino supersymmetry
variation equation boils down to $\Omega_{-\mu}^{ab}\g_{ab}\e_+=0$
which in turn implies that ${R_{\mu\nu}(\Omega_-)}^{ab}\g_{ab}\e_+=0$.
The index-pair interchange symmetry
$R(\Omega_+)_{ab,cd}=R(\Omega_-)_{cd,ab}$, allows us to convert
the previous condition for $\e_+$ to
${R_{\mu\nu}(\Omega_+)}^{ab}\g^{\mu\nu}\e_+=0$.
If we then make the identification
${F_{\mu\nu}}^{ab}\sim {R_{\mu\nu}(\Omega_+)}^{ab}$, we see that
we have reproduced the gaugino supersymmetry variation equation. This
is simply to emphasize that, because of the crucial role of the antisymmetric
tensor in these solutions, the precise way in which the $\Omega_+$'s and
$\Omega_-$'s appear in the various equations we deal with is tightly
constrained and quite critical.

\newsec{Development and Interpretation of the Solutions}

Now we will work out the geometry and physical interpretation of the solutions
described in the previous section.
We begin with a discussion of the ``gauge'' solution described by \stromdil.

There are two charges one can associated with this solution. These are
the instanton winding number
\eqn\instwind{ \nu = {1 \over 480 \pi^2} \int {\rm Tr} F \wedge F ,}
where the integral is over a four-dimensional cross section, and the
axion charge
\eqn\axcharge{Q = - {1 \over 2 \pi^2} \int H , }
where the integral is over an asymptotic $S^3$ sorrounding the fivebrane.
The instanton winding number is quantized as usual and the minimal value
is $\nu=1$.
The axion charge $Q$ is also quantized in integer multiples of $\apm$.
The point is that, if the flux of $H$ through a non-trivial $S^3$ is non-zero,
then
there cannot be a unique two-form potential $B$ covering the whole sphere. The
best one can do is to have two sections, $B^\pm$, covering the upper and
lower halves of the $S_3$ and related to each other by a gauge transformation
in an overlap region which is topologically an $S_2$. Since the sigma model
action involves $B$, not $H$, the non-uniqueness of $B$ could lead to an
ill-defined sigma model path integral. It is possible to
show that, with our definitions of the sigma model action, this danger is
avoided if and only if the flux of $H$ is an integral multiple of $\apm$. The
details of this argument can be found in \wzwpap~.
Although the minimal allowed value of $Q$ is $\apm$, the gauge solution
in fact has $Q=8 \apm$ as can be verified by explicit calculation.

The form of the solution \stromdil\ was determined by a ``low-energy''
expansion
with the expansion parameter being $\apm/ \rho^2$. Solutions of scale size
less than $\sqrt{\apm}$ can be shown to
exist using arguments based on low-energy supersymmetry,
but are not well approximated by the solution of \stromdil.

We now turn to the ``symmetric'' solution.
In the previous section we found that the following is a
solution of the low-energy spacetime effective action of the heterotic string:
\eqn\recap{\eqalign{
	ds^2=&e^{2\phi(x)}\delta_{\mu\nu}dx^\mu dx^\nu+
		\eta_{\alpha\beta}dy^\alpha dy^\beta \cr
            H_{\mu\nu\lambda}=& - {\e_{\mu\nu\lambda}}^\sigma
\p_\sigma \phi\cr
	{F_{\mu\nu}}^{mn}=&\hbox{$\widetilde F_{\mu\nu}$}^{mn}=
                {R_{\mu\nu}(\Omega_-)}^{mn} ~,}}
where $\mu,\nu =6\ldots 9$ , $\alpha,\beta =0\ldots 5$ and $\eta_{\alpha\beta}$
is the Minkowski metric. The last equation expresses the fact that the
gauge field is a self-dual instanton with moduli chosen so that it coincides
(up to gauge transformations of course) with the curvature of the generalized
connection of the theory. The consistency condition for all this is just
$\sq e^{2\phi}=0$.

The solution of the consistency condition on $\phi$ is just a constant
plus a sum of poles:
\eqn\phisoltwo{e^{2\phi}=e^{2\phi_0}+\sum_{i=1}^N{Q_i\over (x-x_i)^2}}
which should be compared with the analogous expression \phisol\ which
appeared in the 'tHooft ansatz.
The constant term is fixed by the (arbitrary) asymptotic value of the dilaton
field, $\phi_0$~. In string theory, $e^{\phi}$ is identified with the local
value of the string loop coupling constant, $g_{str}$. For the solution
described by \phisoltwo\ , $g_{str}$ goes to a constant at spatial infinity and
goes to infinity at the locations of the poles! We shall worry about the
physical interpretation of this fact in due course. Now, the metric of our
solution is conformally flat with conformal factor given by \phisoltwo\ . Since
$\phi$ goes to a constant at infinity, the geometry is asymptotically flat,
which is precisely what we want for a soliton interpretation.
In the neighborhood of a singularity, we can replace
$e^{2\phi}$ by a simple pole $Q/r^2$ and obtain the approximate line element
\eqn\wrmmet{\eqalign{
ds^2\sim & {Q\over r^2}(dr^2 + r^2 d\Omega^2_3) \cr
        =& dt^2 + Q d\Omega^2_3 }}
where $d\Omega^2_3$ is the line element on the unit three-sphere and we have
introduced a new radial coordinate $t=\sqrt{Q}log(r/\sqrt{Q})$. This expression
becomes more and more accurate as $t\to -\infty$.
In this same limit, the other fields are given by
\eqn\wrmfld{\phi = - t/\sqrt{Q} \quad  H= - Q\e_3 ~,}
where $\e_3$ is the volume form on the three sphere. In Sect. 5 we will see
that the linear behavior of the dilaton field plays a crucial role in
the underlying exact conformal field theory. The geometry
described by \wrmmet\ is a cylinder whose cross-section is a three-sphere of
constant area $2\pi^2 Q$. The global geometry is that of a collection of
semi-infinite cylinders, or semi-wormholes (one for each pole in $e^{+2\phi}$),
glued into asymptotically flat four-dimensional space. The semi-wormholes are
semi-infinite since the approximation of \wrmmet\ becomes better and better
as $t\to -\infty$ and breaks down as $t\to +\infty$. It is these semi-wormholes
which we propose to interpret as solitons.

A further crucial fact is that the residues, $Q$, are quantized.
Consider an $S_3$ which surrounds a single pole, of residue Q,
in $e^{+2\phi}$. The net flux of $H$ through this $S_3$ is entirely due
to the enclosed pole and can easily be calculated:
\eqn\Hflux{\eqalign{
\int_{S_3}H=& - 2\pi^2 Q}}
The consequence for us
is that the residues of the poles in $e^{2\phi}$ are discretely quantized:
$Q_i=n_i\apm$. As a result, the cross-sectional areas of the individual
semi-wormholes are quantized in units of $2\pi^2\apm$ and there is thus a
minimal
transverse scale size of the fivebrane. (This fact may be useful in future
attempts to quantize the transverse fluctuations of the fivebrane.)

Finally, we want to characterize the instanton component of this solution.
The key point is that, when the dilaton field satisfies $\sq e^{2\phi}=0$,
we can construct a self-dual $SU(2)$ connection out of the scalar field
$\phi$ and we want to identify this connection with the gauge instanton.

But this is simple in terms of the ansatz \thooftans\ and the corresponding
scalar field \phisol.
An important point is that the total
instanton number of the solution built on $f$ of \phisol\ is $N$, the number of
poles.
The gauge potential which follows from taking $f$ to have a
single pole is
\eqn\instn{ A_\mu=-2\rho^2\bar\Sigma_{\mu\nu}{x^\nu\over x^2(x^2+\rho^2)}~,}
an expression which one immediately recognizes as the singular gauge instanton
of scale size $\rho$ centered at $x=0$. The only way this can match our
construction of a self-dual generalized connection is if we make the
identification
\eqn\ident{f(x)=e^{- 2\phi_0}e^{2\phi}=
                        1+\sum^N_{i=1}{e^{ - 2\phi_0}Q_i\over (x-x_i)^2}~.}
Thus, given the solution \phisoltwo\ for the dilaton field, we can assert that
the associated instanton has instanton number $N$, with instantons of scale
size $\rho_i^2=e^{-2\phi_0}Q_i \apm$ localized at positions $x_i$. Since the
$Q_i$ are
quantized, so are the instanton scale sizes. The only free parameters (moduli)
are the $4N$ center locations of the instantons. In ten dimensions, the
multiple instanton solution corresponds to multiple fivebranes with the
locations in the transverse four-dimensional space of the individual fivebranes
given by the center coordinates of the individual instantons.

An important fact about the solution we have just constructed is that it is
not perturbative in $\apm$. As we saw in the discussion following \recap ,
the semi-wormhole associated with a pole of residue $Q=n\apm$ has a
cross-section
which is a sphere of area $2\pi^2Q$ and therefore has curvature $R\sim 1/Q$.
In the perturbative sigma model approach to strings in background fields,
one finds that the sigma model expansion parameter is $\apm R$. In the case
at hand, this becomes $\apm/Q= 1/n$, which is obviously not small for the
elementary fivebrane, which has $n=1$. Since our solution has been constructed
by solving the leading-order-in-$\apm$ beta function equations, ignoring all
higher-order corrections, one can legitimately worry whether it makes any
sense. In the next two sections we will present evidence that all
corrections to this particular solution actually vanish, and the leading-order
solution is exact.

There is another perturbation theory issue which should be mentioned
here. String theory has
two expansion parameters: the string tension $\apm$ and the string loop
coupling constant $g_{str}\sim e^{\phi_0}$. The latter is the quantum
expansion parameter of string theory and, in this paper, we are working to
zeroth-order in an expansion in $g_{str}$. In effect, we are producing an
exact solution of {\it classical} string field
theory. However, as we have already pointed out, our solution has the unusual
feature that $g_{str}$ grows without limit down the throat of a semi-wormhole
so that there is, strictly speaking, no reliable classical limit! Since
virtually nothing is known about non-perturbative-in-$g_{str}$ physics, we
don't know what this means for the ultimate validity of this sort of solution.
Similar issues arise in the matrix model/Liouville theory approach to
two-dimensional quantum gravity, and we hope eventually to gain some insight
from that source.

Another interesting point concerns what happens when we lift the requirement
of spacetime supersymmetry and look for solutions of the beta function
equations rather than the condition that some supersymmetry charges annihilate
the solution. Our solutions have the property that the mass (the ADM mass, to
be precise) per unit fivebrane area is proportional to the axion charge:
$M_5=Q$. This equality can be understood via a Bogomolnyi
bound: any solution of the leading-order field equations with the fivebrane
topology must satisfy the inequality $M_5\ge Q$ and our
solution saturates the inequality. One can easily imagine a process in which
mass, but not axion charge, is increased by sending a dilaton wave down one of
the semi-wormhole throats. Since the semi-wormhole throat is semi-infinite,
this wave
need
not be reflected back: It can continue to propagate down the throat forever,
leaving an exterior solution for which $M_5>Q$. Such solutions
of the leading-order beta function equations have indeed been found \bhole\
and they resemble the familiar Reissner-Nordstrom family of charged black
holes: they have an event horizon and a singularity, but the singularity
retreats to infinity as the mass is decreased to the extremal value that
saturates the Bogomolny bound. Perhaps not surprisingly, the non-extremal
solutions are not annihilated by any spacetime supersymmetries.
Nevertheless it is possible in some cases to give exact conformal field
theory constructions of these solutions \GiStr.
These developments
should eventually allow us to make progress on understanding the
string physics of black holes, Hawking radiation and the like.

In the rest of these lectures, we will pursue the much more limited goal of
showing that the symmetric solution is an exact solution of string theory
using world-sheet arguments.

\newsec{Worldsheet Sigma Model Approach}

To show conclusively that a given spacetime configuration is a solution of
string theory, we must show that it derives from an appropriate
superconformal worldsheet sigma model. In this section we will show that
the worldsheet sigma models corresponding to the fivebranes constructed
in section 5 possess extended worldsheet supersymmetry of type (4,4)
The notation derives from
the fact that in a conformal field theory, the left-moving fields (functions
of $z$) and the right-moving fields (functions of $\bar z$) are dynamically
independent. It is therefore possible to have different numbers of right- and
left-moving supercharges $Q^I_\pm$. The general case, referred to as
$(p,q)$ supersymmetry, is described by the algebra
\eqn\extalg{\eqalign{
\{ Q^I_+, Q^J_+\}=& 2 \delta^{IJ} P_+ \qquad I,J=1\ldots p\cr
\{ Q^I_-, Q^J_-\}=& 2 \delta^{IJ} P_- \qquad I,J=1\ldots q\cr
\{ Q^I_+, Q^J_-\}=&0~~.}}
The minimal possibility, corresponding to a generic solution of the heterotic
string, has $(1,0)$ supersymmetry. Any left-right-symmetric, and therefore
non-anomalous theory, will have $(p,p)$ supersymmetry (this is sometimes
referred to as $N=p$ supersymmetry). The maximal possibility
is $(4,4)$ which, it turns out, is what is realized in our fivebrane solution.
We will argue that, in the (4,4) case,
there is a nonrenormalization theorem which makes the lowest-order in $\apm$
solution for the spacetime fields exact. The latter issue is closely related
to the question of finiteness of sigma models with torsion and with extended
supersymmetry \refs{\howpaptwo, \vnew} and the results we find are slightly
at variance with the conventional wisdom, at least as we understand it. We will
comment upon this at the appropriate point.

First we digress to explain why we expect four-fold extended
supersymmetry in this problem. The models of interest to us are structurally
equivalent to a compactification of ten-dimensional spacetime down to six
dimensions: there are six flat dimensions (along the fivebrane) described
by a free field theory and four `compactified' dimensions (transverse to the
fivebrane) described by a nontrivial field theory. The fact that the
`compactified' space is not really compact has no bearing on the supersymmetry
issue. The defining property of all the fivebranes of section 5 is that they
are
annihilated by the generators of a six-dimensional N=1 {\it spacetime}
supersymmetry. That is, they provide a compactification to six
dimensions which maintains N=1 spacetime supersymmetry. Now, it is well-known
that in compactifications to four dimensions, the sigma model describing the
six compactified dimensions must possess (2,0) worldsheet supersymmetry in
order for the theory to possess N=1 four-dimensional spacetime supersymmetry
\BDHM. Roughly speaking, the conserved U(1) current of the (2,0)
superconformal algebra defines a free boson which is used to construct the
spacetime supersymmetry charges. It is also known that, if one wants to impose
N=2 four-dimensional spacetime supersymmetry, the compactification sigma model
must have (4,0) supersymmetry \BD. The conserved SU(2) currents of the (4,0)
superconformal algebra are precisely what are needed to construct the larger
set
of N=2 spacetime supersymmetry charges. Since, by dimensional reduction, N=1
supersymmetry in six dimensions is equivalent to N=2 in four dimensions, the
above line of argument implies that spacetime supersymmetric compactifications
to six dimensions (including our fivebrane) require a compactification sigma
model with at least (4,0) worldsheet supersymmetry. Since our solution is
constructed to cancel the anomaly, it will be left-right symmetric and
therefore automatically of type $(4,4)$.

Now we turn to a study of string sigma models. The generic
sigma model underlying the heterotic string describes the dynamics of D
worldsheet bosons $X^M$ and D right-moving worldsheet fermions $\psi^M_R$
(where D, typically ten, is the dimension of spacetime) plus left-moving
worldsheet fermions $\lambda^a_L$  which lie in a representation of the gauge
group $G$ (typically $SO(32)$ or $E_8\otimes E_8$). The generic Lagrangian for
this sigma model is written in terms of coupling functions $G_{MN}$,
$B_{MN}$ and $A_M$ which eventually get interpreted as spacetime metric,
antisymmetric tensor and Yang-Mills gauge fields. This Lagrangian  has
the explicit form \hetsigref\
\eqn\hetsiglag{\eqalign{{1\over
4\pi\apm}\int d^2\sigma\{ &
       G_{MN}(X)\partial_+X^M\partial_-X^N +
            2   B_{MN}(X)\partial_+X^M\partial_-X^N \cr
+&iG_{MN}\psi^M_R{\cal D}_-\psi^N_R +i\delta_{ab}\lambda^a_L{\cal D}_+
\lambda^b_L
     +\half (F_{MN})_{ab}\psi^M_R\psi^N_R\lambda^a_L\lambda^b_L\}~~}}
where $H=dB$.  In this expression, the covariant derivatives on the
left-moving fermions are defined in terms of the Yang-Mills connection,
while the covariant derivatives on the right-moving fermions are defined in
terms of a non-riemannian connection involving the torsion (which already
appeared in section 5):
\eqn\covderiv{\eqalign{
{\cal D}_- \psi^A_R=&\partial_-\psi^A_R+
    {{\Omega_{-N}}^A}_B \partial_-X^N\psi^B_R, \cr
{\cal D}_+\lambda^a_L=&\partial_+\lambda^a_L+
    {{A_{N}}^a}_b\partial_+X^N\lambda^b_L . }}
We use indices of type $M$ for coordinate space indices,
type $A$ for the tangent space and type $a$ for the gauge group.
An absolutely crucial feature of this action is that the connection appearing
in
the covariant derivative of the right-moving fermions is the generalized
connection $\Omega_-$, {\it not} the Christoffel connection. This action has
a naive $(1,0)$ worldsheet supersymmetry and can be written in terms of (1,0)
superfields. Superconformal invariance is broken by anomalies of various kinds
unless the coupling functions satisfy certain `beta function' conditions
\cfmpetal\ which are equivalent to the spacetime field equations discussed in
section 5.  The dilaton enters these equations in a rather roundabout, but
by now well-understood, way \hetsigref.

To proceed further, we must construct the specific sigma models corresponding
to the fivebrane solutions. For the generic fivebrane, \hetsiglag\
undergoes a split into a nontrivial four-dimensional theory and a
free six-dimensional theory: the sigma model metric (as opposed to the
canonical general relativity metric) then describes a flat six-dimensional
spacetime times four curved dimensions. The right-moving fermions couple
via the kinetic term to the generalized connection $\Omega_-$, which acts only
on the four right-movers lying in the tangent space orthogonal to the
fivebrane. The other six right-movers are free (we momentarily ignore the
four-fermi coupling) so there is a six-four split of the right-movers as well.
The left-moving fermions couple to an instanton gauge field which
may or may not be identified with the {\it other} generalized
connection, $\Omega_+$. In all the cases of interest to us, the gauge
connection
is an instanton connection and acts only in some $SU(2)$ subgroup of the full
gauge group, so that four of the left-movers couple nontrivially, while the
other 28 are free. Finally, the four-fermion interaction term couples together
precisely those left- and right-movers which couple to the nontrivial gauge and
$\Omega_-$ connections and is therefore consistent with the six-four split
defined by the kinetic terms. The remaining variables can be regarded as
defining a heterotic, but free, theory (6 $X$, 6 $\psi_R$ and 28 $\lambda_L$)
living in the six `uncompactified' dimensions along the fivebrane. From now
on, we focus our attention on the nontrivial piece of \hetsiglag\ referring to
the four-dimensional part of the split. For string theory consistency, it must
have a central charge of 6, which would be trivially true if the connections
were all flat, but is far from obvious for a fivebrane.

Now let us further specialize to the sigma model underlying the left-right
symmetric (and therefore non-anomalous) fivebrane solution of section 5. It is
constructed by identifying the gauge connection with the `other' generalized
connection $\Omega_+$ and making that connection self-dual by imposing the
condition $\sq e^{2\phi}=0$ on the metric conformal factor. The result of
this is that the four bosonic coordinates transverse to the fivebrane and
the four nontrivially-coupled left- and right-moving fermions are governed
by the worldsheet action
\eqn\torsig{\eqalign{{1\over 4\pi\apm}\int
d^2\sigma\{ &
       G_{\mu\nu}(X)\partial_+X^\mu\partial_-X^\nu +
             2  B_{\mu\nu}(X)\partial_+X^\mu \partial_-X^\nu \cr
+&iG_{\mu \nu}\psi^\mu_R{\cal D}_-\psi^\nu_R +iG_{\mu \nu}
\lambda^\mu_L{\cal D}_+\lambda^\nu_L
       +\half R(\Omega_+)_{\mu
\nu \lambda \rho}\psi^\mu_R\psi^\nu_R\lambda^\lambda_L\lambda^\rho_L\} }}
where ${\cal D}_\pm$ are the covariant derivatives built out
of the generalized connections $\Omega_\pm$ . In fact, as long as the $H$
appearing in $\Omega_\pm$ is given by $d B$, \torsig\ is identical to
the basic left-right symmetric, $(1,1)$ supersymmetric nonlinear sigma
model with torsion \hetsigref\ .
Despite the apparent asymmetry of the coupling of $\lambda_L$ to $\Omega_+$
and $\psi_R$ to $\Omega_-$, the theory nonetheless has an overall left-right
symmetry (under which $B \rightarrow -B$) and is non-anomalous. To exchange the
roles of $\psi_R$ and $\lambda_L$ one has to replace the curvature of
$\Omega_-$ by that of $\Omega_+$. This exchange symmetry property relies on
the non-riemannian relation
\eqn\nonriem{ R(\Omega_+)_{\mu \nu \lambda \rho}=
 R(\Omega_-)_{\lambda \rho \mu \nu}}
which indeed holds for the generalized connection \connect\  when $d
H=0$.
To summarize, we have shown that the heterotic sigma model describing the
nontrivial four-dimensional geometry of the fivebrane is actually an example
of a left-right symmetric sigma model with at least $(1,1)$ supersymmetry.
As we will now show, it actually has $(4,4)$ worldsheet supersymmetry.

We now turn to the question of extended supersymmetry. The basic worldsheet
supersymmetry of a $(1,1)$ model like \torsig\ is
\eqn\susyone{ \eqalign{
\d X^M=\e_L\psi^M_R+&\e_R\psi^M_L \cr
\d \psi^A_L+{\Omega_{+M}}^A_{~~B}\d X^M\psi^B_L=&\p X^A\e_R+\ldots \cr
\d \psi^A_R+{\Omega_{-M}}^A_{~~B}\d X^M\psi^B_R=&\p X^A\e_L+\ldots }~.}
The worldsheet supersymmetry of the $(1,0)$
model is obtained by dropping the contributions of $\e_R$ and $\psi_L$.
The general structure of a possible second supersymmetry transformation is
\eqn\susytwo{ \eqalign{
\hat\d X^M=\e_Lf_R(X)^M_{~N}\psi^N_R+&\e_Rf_L(X)^M_{~N}\psi^N_L \cr
\hat\d \psi^A_L+{\Omega_{+M}}^A_{~~B}\d X^M\psi^B_L=&
		-f_L(X)^A_{~B}\p X^B\e_R+\ldots \cr
\hat\d \psi^A_R+{\Omega_{-M}}^A_{~~B}\d X^M\psi^B_R=&
		-f_R(X)^A_{~B}\p X^B\e_L+\ldots ~.}}
The function $f$ is normalized and fully defined by the requirements that
$\{\hat\d,\d\}=0$ and that $\hat\d$ anticommute with itself to give ordinary
translations as in \extalg. The question is, what conditions must $f$
satisfy in order for $\hat\d$ to be a symmetry and how many can
exist?

This question was first addressed in \alfr\ for the case of left-right
symmetric theories without torsion (\ie without an antisymmetric tensor
coupling term). The more complex case of left-right symmetry with torsion
was subsequently dealt with in \refs{\ghr,\howpaptwo,\vnew}. The basic result
is that the pair of tensors $f_{R,L}$ must be complex structures, covariantly
constant with respect to the appropriate connection:
\eqn\cmpstr{\eqalign{
f^2_\pm = & -1 \cr
	{\cal D}^\pm_A {f_\pm}^B_{~C}=& \p_A{f_\pm}^B_{~C}+
		{\Omega^{(\pm)}_{AD}}^B {f_\pm}^D_{~C}-
			{\Omega^{(\pm)}_{AC}}^D {f_\pm}^B_{~D}=0~,}}
where the $\pm$ notation is equivalent to the $L,R$ notation. The tensors in
\cmpstr\ are written in tangent space indices which is why the generalized
spin connections $\Omega^{(\pm)}$ appear in the covariant derivative. The
equation could, of course, also have been written in coordinate indices.
In general, it is not obvious that such a pair of complex structures can be
found, but, if one can, we know that the sigma model actually possesses
$(2,2)$ worldsheet supersymmetry. A further question is whether multiple
pairs $f^{(r)}_\pm$ of such complex structures can be found. If we can find
$p-1$ of them, then the sigma model has $(p,p)$ supersymmetry. It turns out
that the only consistent possibility for multiple complex structures is that
there be three of them \alfr\ and that they satisfy the Clifford algebra
\eqn\clifford{f^{(r)}_\pm f^{(s)}_\pm =-\d_{rs}+\e_{rst}f^{(t)}_\pm ~.}
This corresponds to the case of $(4,4)$ supersymmetry. It is worth noting
that each complex structure leads to a conserved (chiral) current:
\eqn\concur{J_\pm^{(r)}=\psi^A_\pm (f^{(r)}_\pm)_{AB}\psi^B_\pm~.}
This yields a $U(1)$ symmetry in the $(2,2)$ case and an $SU(2)$ symmetry in
the $(4,4)$ case.

The question of left-right asymmetric theories, such as those which underlie
the ``gauge'' fivebranes discussed in section 5, is more delicate.
According to \howpaptwo\ , a heterotic sigma model will have $(p,0)$
supersymmetry if there are $p-1$ complex structures $f^{(r)}_+$ which
are covariantly constant under the connection which couples to the right-moving
fermions (those which do not couple to the gauge field) and if the gauge field
(which affects the left-moving fermions) satisfies a condition which reduces,
for a four-dimensional base space, to self-duality. The latter condition
is met for all of the fivebranes of interest to us since they are all built
on instanton gauge fields. Thus, in all cases, the essential issue is the
existence of complex structures.

To count complex structures, we will use the connection between complex
structures and covariantly constant spinors (a nice pedagogical discussion
can be found in \candelas).
We start with a spinor $\eta$ (in our case four-dimensional) of definite
chirality ($\g_5\eta=+ \eta$, say) and unit normalized ($\eta^\dagger\eta=1$).
Then we define a tensor
\eqn\construct{J_{AB}=-i\eta^\dagger\g_{AB}\eta}
which we will
identify as a complex structure tensor (in tangent space indices and with
indices raised and lowered by the identity metric). It is then
automatic that if the spinor is covariantly constant with respect
to some connection, so is $J_{AB}$. A simple Fierz identity argument
then shows that $J$ squares
to $-1$ (${J_A}^B{J_B}^C=-{\d_A}^C$) and is indeed a complex structure.

We are now ready to construct the explicit complex structures.
As was explained in the discussion after \sutwo\ , on the fivebrane,
{\it constant} spinors of definite four-dimensional chirality are covariantly
constant. Using the Weyl representation for the four-dimensional gamma
matrices, one has the following solutions of the two covariant constancy
conditions:
\eqn\cstspin{\eqalign{
\cD_\mu(\Omega_+)\e_+=0~&\Rightarrow~\e_+=\pmatrix{\chi\cr 0} \cr
\cD_\mu(\Omega_-)\e_-=0~&\Rightarrow~\e_-=\pmatrix{0\cr \chi} ~,}}
where $\chi$ is {\it any} constant two-spinor (which we might as well unit
normalize). Since there are three parameters needed to specify the general
normalized two-spinor, there should be three independent choices for the
two-spinor $\chi$ and therefore three choices for both $\e_+$ and $\e_-$.
We will define the independent $\chi_r$ ($r=1,2,3$) as those which give
expectation values of the spin operator along the three coordinate axes:
\eqn\expect{ \chi^\dagger_r\sigma^i\chi_r=\d_{ir} .}
This finally leads, with the help of \construct\ , to the following set of
three right- and left-handed complex structures:
\eqn\result{\eqalign{
J^+_1=\pmatrix{i\s_2&0\cr 0&i\s_2} &\quad
										J^-_1=\pmatrix{-i\s_2&0\cr 0&-i\s_2}\cr
J^+_2=\pmatrix{0&1\cr -1&0} &\quad
										J^-_2=\pmatrix{0&-\s_3\cr \s_3&0}\cr
J^+_3=\pmatrix{0&i\s_2\cr i\s_2&0} &\quad
										J^-_3=\pmatrix{0&-\s_1\cr \s_1&0}~.}}
It is trivial to show that the $J^+$ commute with all the $J^-$ and that they
satisfy the Clifford algebra \clifford . These are precisely the conditions
needed to generate $(4,4)$ supersymmetry in a left-right symmetric theory
(or $(4,0)$ supersymmetry in a heterotic theory). The complex structures are
thus extremely simple indeed.

Finally, we come to the questions of finiteness and need for higher-order
or non-perturbative
in $\apm$ corrections to our solutions. It is rather firmly established
that two-dimensional nonlinear sigma models with $(4,4)$ supersymmetry
{\ without} torsion ($B_{\mu\nu}=0$) are
in fact finite. The general proof was given quite some time ago by
Alvarez-Gaume
and Freedman \alfr\ and assumes that $(4,4)$ supersymmetry is not
explicitly broken at the quantum level. They then show that
no $(4,4)$-invariant
counterterms - perturbative or non-perturbative-
of the needed dimension can be constructed. If the theory is
finite, the beta-functions get no higher-order corrections and the choice of
background fields which made the beta functions vanish at leading order must
continue to make them vanish at all orders in $\apm$.
A similar result was shown in \wsheet\ to hold, relying heavily on the
results of \refs{\ghr,\howpaptwo, \vnew} for $(4,4)$ models with torsion.
(These arguments are backed up at the perturbative level by
superfield non-renormalization theorems\howpaptwo.) The functional
form of the action must satisfy certain conditions in order to have $(4,4)$
supersymmetry and one can see that the most general solution
of these conditions corresponds precisely to our special multi-fivebrane
solution.

As an aside, we mention that it has been argued that one really only needs
$(4,0)$ supersymmetry to achieve finiteness \howpaptwo . This would apply
to variations on the solution described in Sect.~2 in which, for example,
the gauge instanton
scale size did not match the semi-wormhole throat transverse
scale size or to the original instanton solution\stromdil.
In the discussion given earlier in this section,
we recall that the existence and properties of the right-moving complex
structures $f^{(+)}_i$ have nothing to do with the properties of the gauge
field (which governs the left-moving complex structures). So, if we keep the
same metric then we should have the same $f^{(+)}_i$ and thus at least a
$(4,0)$ supersymmetry. In this case
there will be corrections to the beta functions so that
the theory is not finite, but may be constructible order by order, as was
shown in \wbran\ for the solution \stromdil\ using spacetime methods.
This subject has
yet to be explored in any detail from the worldsheet point of view.

\newsec{Algebraic CFT Approach}

It is one thing to show that a sigma model is a superconformal field theory, as
we have done in the previous section, and quite another to be able to classify
its primary field content and calculate n-point functions of its vertex
operators. Indeed, in order to answer all the interesting questions about
string
solitons, it would be desirable to have as detailed an algebraic
understanding of the underlying conformal field theory as we already have for,
say, the minimal models. We are far from having such an understanding,
but in this section we will see that in some case
useful progress can be made.

Recall from section 6 that the (four-dimensional part of the) metric
of the symmetric solution has the form
\eqn\mmetric{ds^2=e^{2\phi}dx^2}
where $dx^2$ is the flat metric on $R^4$ and
\eqn\confact{e^{2\phi(x)}=e^{2\phi_0}+\sum_1^n{Q_i\over (x-x_i)^2}~.}
The singularities in $e^{2\phi}$ are associated with the semi-wormholes.
Taking $n=1$ and the limit $e^{2\phi_0} \rightarrow 0$ gives
\eqn\wormlim{e^{2\phi}= {Q\over x^2},}
which is the solution corresponding to the semi-wormhole throat
itself. Using spherical coordinates
centered on the singularity, and defining a logarithmic radial coordinate by
$t=\sqrt{Q}ln \sqrt{x^2/Q}$, the metric, dilaton  and axion
field strength of the throat may be written in the form
\eqn\wormsol{\eqalign{ds^2&= dt^2 + Q d\Omega_3^2, \cr \phi &
=-t  / \sqrt{Q}, \cr H &=-Q\epsilon,} }
where $d\Omega_3^2$ is the line element and $\epsilon$ the volume form
of the unit 3-sphere obeying $\int \epsilon =2\pi^2$. The geometry of the
throat is thus a 3-sphere of radius $\sqrt{Q}$ times the open line $R^1$
and the dilaton is linear in the coordinate of the $R^1$.
Remarkably, these metric and antisymmetric tensor fields are such that the
curvatures constructed from the generalized connections, defined in \connect\ ,
are identically zero, reflecting the parallelizability of $S^3$.
The axion charge $Q$ is integrally quantized. So, since $Q$ appears in the
metric, the radius of the $S^3$ is quantized as well.

The sigma model defined by these background fields is an interesting variant of
the Wess-Zumino-Witten model and the underlying conformal field theory can, it
turns out, be analyzed in complete detail. The basic observation along these
lines was made in \bachas\ in the lorentzian context and euclideanized in
\refs{\hrs, \sjrey}: the $S^3$ and the antisymmetric tensor field are
equivalent to the $O(3)$ Wess-Zumino-Witten model of level
\eqn\level{k={Q\over \apm} ,} while the $R^1$ and the linear
dilaton define a Feigin-Fuks-like free field theory with a background charge
induced by the linear dilaton. Both systems are conformal field
theories of known central charges:
\eqn\wormccs{c_{wzw}={3k\over k+2}\qquad c_{ff}=1+{6\over k}~~.}
The shift of the $R^1$ central charge away from unity is a familiar background
charge effect which has been exploited in constructions of the minimal
models \rusref\ and in cosmological solutions \bachas.

For the combined
theory to make sense, the net central charge must be four. Let us for the
the moment consider the bosonic string. If we expand $c_{wzw}$ in powers of
$k^{-1}$ (this corresponds to the usual perturbative expansion in powers of
$\apm$), we see instead that
\eqn\ctot{c_{tot}=c_{wzw}+c_{ff}=4 + O(k^{-2}\sim{\apm}^2) ~~.}
But, we should not have expected to do any better: the
field equations we solved in section 5 to get this solution are
only the leading order in $\apm$ approximation to the full bosonic string
theory field equations and we must expect higher-order corrections to the
fields
and central charges. In fact, this issue can be studied in detail and it
can be shown \ramzi\ that the metric and antisymmetric tensor fields are not
modified and that the only modification of the dilaton is to adjust the
background charge of the $R^1$ ({\it i.e.} the coefficient of the
linear term in $\phi$) so as to maintain $c_{tot}$ exactly equal to four.

While this is quite interesting, we are really interested in the
superstring case. The leading-order-in-$\apm$ metric, dilaton etc.
fields are the same as in the bosonic case (and, because of the
non-renormalization theorems, we
expect no corrections to them) but various fermionic terms are added to the
previous purely bosonic sigma model. The structure is that of the (1,1)
worldsheet supersymmetric sigma model \torsig\ discussed in section 7. There
is still an $S^3\times R^1$ split, but the component theories are
supersymmetrized versions of Wess-Zumino-Witten and Feigin-Fuks. The
Feigin-Fuks theory is still essentially free. In the supersymmetric WZW theory,
the four-fermi terms vanish identically because, as pointed out above,
the generalized curvature vanishes for this background. As a consequence, the
generalized connections are locally pure gauge
and can be eliminated from the fermion
kinetic terms by a gauge rotation of the frame field.
Since the fermions are effectively free, they
make a trivial addition to the  central charges of both the $S^3$ and the
$R^1$ models:
\eqn\fwormcc{c_{wzw}={3 k\over k+2}+{3\over 2}
                             \qquad c_{ff}=1+{6\over k}+{1\over 2}~~.}
There is, however, a small subtlety: the gauge rotation which decouples the
fermions is {\it chiral}, and therefore anomalous, because the left- and
right-moving fermions couple to two {\it different} pure gauge generalized
connections, $\Omega_+$ and $\Omega_-$. The entire effect of this anomaly on
the
central charge turns out to be the replacement in $c_{wzw}$ of $k$ by $k-2$
(the details can be found in \ryanetal) with the result that
\eqn\anomcc{c_{wzw}={3 (k-2)\over k}+{3\over 2}
					\qquad c_{tot}=c_{wzw}+c_{ff}=6~~.}
Six is, of course, exactly the value we want for the central charge. The
remarkable fact is that, in the supersymmetric theory,
the expansion of $c_{wzw}$
in powers of $k^{-1}$ {\it terminates} at first non-trivial order and no
modification of the dilaton field is needed to maintain the desired central
charge of six. These results are consistent with the non-renormalization
theorems discussed in section 7, but are not tied to perturbation theory,
since they derive from exactly-solved conformal field theories. On the other
hand, since the present discussion makes no reference to the (4,4)
supersymmetry which was crucial in proving the perturbative
non-renormalization theorems of section 7, an important element
is still missing.

This is a good point to remind the reader of the hierarchy of superconformal
algebras. Much of what we know about conformal field theory comes from studying
the representation theory of these algebras. The basic N=1 superconformal
algebra is contains an energy-momentum tensor $T(z)$ and its superpartner
$G(z)$.
The essential information is contained in the
algebra obeyed by their Laurent coefficients $L_n$ and $G_r$:
\eqn\algone{\eqalign{
[L_m,L_n] =& {\hat c \over 8} m (m^2 -1) \delta_{m+n,0} + (m-n) L_{m+n} , \cr
\{ G_r,G_s\} =& {\hat c \over 2} (r^2 -{1 \over 4}) \delta_{r+s,0} + 2 L_{r+s}
, \cr
[L_m,G_r] =& ({m \over 2} -r) G_{m+r}  }}
with $\hat c = 2c/3$ in terms of the usual conformal anomaly.
%
All superstring theories have at
least this much worldsheet supersymmetry. The N=2 superconformal algebras
differ from this by having a conserved current $J(z)$ and {\it two}
supercharges $G^\pm(z)$ distinguished by the value ($\pm 1$) of their charge
with respect to the current $J(z)$. This charge also plays a key role in the
GSO projection which rids the theory of tachyons. The important new algebraic
relations are contained in the
commutation relations
\eqn\algtwo{\eqalign{
[J_m,G_r^\pm] =& \pm G_r^\pm , \cr
\{G_r^+,G_s^+\}=&\{G_r^-,G_s^-\}=0 , \cr
\{G_r^+,G_s^-\} =& L_{r+s} + {1 \over 2} (r-s) J_{r+s} +
               {\hat c \over 4} (r^2 -{1 \over 4}) \delta_{r+s} }}
%
%
%
There is an N=1 subalgebra generated by $T(z)$ and
$G(z)={1\over\sqrt{2}}(G^+(z)+G^-(z))$. The `practical' utility of the N=2
algebra is that the conserved current defines a free field $H$ by the relation
$J(z)=i\sqrt{c\over 3}\p_z H(z)$ and this free field can be used to construct
the N=1 {\it spacetime} supersymmetry charge in a compactification to four
dimensions \BDHM.
One further extension, to four supercharges, turns out to be possible.
There are
now three conserved currents $J^i$ which generate an $SU(2)$ Kac-Moody algebra
and the supercharges $G^\a(z),\bar G^\a(z)$ are in $I=1/2$ representations
of the conserved $SU(2)$.
%
%
The relevant (anti-) commutation relations are
\eqn\algfour{\eqalign{
[L_m,L_n]=& (m-n)L_{m+n} + {k \over 2} m (m^2-1) \delta_{m+n,0} , \cr
\{G_r^\alpha , G_s^\beta \} =& \{\bar G_r^\alpha , \bar G_s^\beta \} =0 , \cr
\{G_r^\alpha , G_s^\beta \} =& 2 \delta^{\alpha \beta} L_{r+s} -
                            2(r-s) \sigma_{\alpha \beta}^i J_{r+s}^i +
                           {k \over 2} (4 r^2 -1) \delta_{r+s,0} , \cr
[J_m^i , J_n^j ] =& i \epsilon^{ijk} J_{m+n}^k + {1 \over 2} k m \delta_{m+n,0}
                   \delta_{ij} , \cr
[J_m^i , G_r^\alpha ] =& - \half \sigma_{\alpha \beta}^i G_{m+r}^\beta , \cr
[J_m^i , \bar G_r^\alpha ] =& \half \bar \sigma_{\alpha \beta}^i \bar
G_{m+r}^\beta }}
with $\sigma^i$ the usual Pauli matrices and $\bar \sigma^i$ their complex
conjugates.

The triplet of conserved charges is what is needed to construct the larger
spacetime supersymmetry algebra associated with a compactification down to
six, rather that four, dimensions. The $SU(2)$ Kac-Moody algebra is of
arbitrary level $k$, but we can see by comparison with \algone\ that the
central charge $c$ is constrained to be $6k$. Since
the level is constrained by unitarity to be integer, the only allowed values
of the central charge are $6,12,\ldots$~. Fortunately, $c=6$ is just what we
need, and this suggests that the N=4 algebra will be important for us.

We will now show that a closer examination of the algebraic structure of the
throat conformal field theory reveals the existence of just the right
extended
supersymmetry. An important clue to understanding the structure of the
$(4,4)$ superconformal symmetry comes from the fact that there must be
{\it two} $SU(2)$ Kac-Moody symmetries:
The first is part of the standard N=4 superalgebra. This algebra contains the
energy-momentum tensor $T(z)$, four supercurrents $G^a(z)$ and three currents
$J^i(z)$ of conformal weight 1, which generate an $SU(2)$ Kac-Moody algebra
of a level tied to the conformal anomaly (in our case, level one).
The second is the $SU(2)$ Kac-Moody algebra of the Wess-Zumino-Witten
part of the throat conformal field theory. It has a general level $n$,
related to the area of the throat cross-section (or, equivalently, its
axion charge) and is clearly distinct from the N=4 $SU(2)$ Kac-Moody.
Since the superconformal algebra is quite tightly constrained, it is not
{\it a priori} obvious that such an $SU(2) \otimes SU(2)$ Kac-Moody is
compatible with N=4 supersymmetry and useful information, such as restrictions
on allowed values of the central charge, might be obtained by explicitly
constructing the algebra (assuming a consistent one to exist).
Quite remarkably, precisely the algebra we need has already been constructed by
Schoutens in \schout. A closely related version of this algebra
was presented in
Sevrin et. al. \sevrin\ , as an alternate N=4 superalgebra,
containing an $SU(2)\otimes SU(2) \otimes U(1)$ Kac-Moody algebra, which had
been missed in previous attempts at a general classification of extended
superalgebras. Similar results also appear
in \ivan. In what follows\footnote*{This discussion was developed
in collaboration with E. Martinec.}
we briefly summarize enough of this work to explain its significance for the
throat problem and, in particular, to verify the assertions made in section 7
about the r\^ole of (4,4) supersymmetry. In addition to establishing the
presence of a $(4,4)$ superconformal symmetry, this construction is a useful
starting point for studying the correspondence between the instanton moduli
space and perturbations of the superconformal field theory.

The construction discovered by Sevrin et.al. goes as follows: Start with the
bosonic WZW model for an $SU(2)\otimes U(1)$ group manifold (this is the
geometry of the throat if we let the radius of the $U(1)$ be infinite). The
conformal model contains four dimension-one Kac-Moody currents, $J^i$
$i=1,2,3$ which satisfy the Kac-Moody algebra of $SU(2)$ with level $n$ and
an additional $U(1)$ current $J^0$ which satisfies the $U(1)$ algebra with
level one.
This is supersymmetrized by adding
a
set of four dimension-1/2 fields $\psi^a$ satisfying the {\it free} fermion
algebra
(this is motivated by the arguments given earlier in this section that the
fermions in a supersymmetric WZW model are, modulo anomalies, free).

As usual, the Sugawara construction provides an energy-momentum tensor
\eqn\sugawar{T(z)=-J^0J^0-{1\over n+2}J^iJ^i-\partial\psi^a\psi^a}
with respect to which the fields $J^a$ ($\psi^a$) are primaries of weight
1 (1/2) and which has the expected $S_{wzw}$  conformal anomaly
\eqn\cexpct{c_{swzw}={3n\over n+2}+3=6(n+1)/(n+2)~.}
There is also a Sugawara-like construction of four real supersymmetry charges
$G^a$, with $a=0,..,3$ :
\eqn\suchrg{\eqalign{
G^0=&2[J^0\psi^0+(1/\sqrt{n+2})J^i\psi^i+(2/\sqrt{n+2})\psi^1\psi^2\psi^3] \cr
G^1=&2[J^0\psi^1+(1/\sqrt{n+2})(-J^1\psi^0+J^2\psi^3-J^3\psi^2)-
				(2/\sqrt{n+2})\psi^0\psi^2\psi^3]\cr}}
(plus cyclic expressions for $G^2$ and $G^3$). These supercharges could have
been packaged as a complex $I=1/2$ multiplet, as in \algfour. The operator
product expansion of these supercharges with themselves reads
\eqn\gope{\eqalign{
G^a(z) G^b(w)=&4{(n+1)\over(n+2)}\delta^{ab}(z-w)^{-3}+
            2\delta^{ab}T(w)(z-w)^{-1} \cr
-8[&{1\over n+2}\alpha^{+i}_{ab}A^+_i(w)+
           {n+1\over n+2}\alpha^{-i}_{ab}A^-_i(w)](z-w)^{-2}\cr
-4[&{1\over n+2}\alpha^{+i}_{ab}\partial A^+_i(w)+
           {n+1\over n+2}\alpha^{-i}_{ab}\partial A^-_i(w)](z-w)^{-1}\cr}}
where
\eqn\thooft{\alpha^{\pm i}_{ab}=\pm\delta^i_{[a}\delta^0_{b]}
           +\half\epsilon_{iab} }
and
\eqn\sutoo{A^-_i=\psi^0\psi^i+\epsilon_{ijk}\psi^j\psi^k\qquad
A^+_i=-\psi^0\psi^i+\epsilon_{ijk}\psi^j\psi^k +J^i}
are commuting $SU(2)$ Kac-Moody algebras of levels 1 and n+1, respectively.
The c-number term (the central charge) and the term involving $T(z)$ are
obligatory in any higher-N superalgebra, while the terms involving dimension 1
operators are what differentiate the various possible extended superalgebras.
With further effort, one shows that the $G\cdot A^\pm$ OPE generates
combinations of $G^a$ and $\psi^a$ while the $G\cdot \psi$ OPE yields $A^\pm$
and $J^0$. No new operators appear in further iterations, so the complete
algebra generated by the supercharges contains just $T$ (dimension 2), $G^a$
(dimension 3/2), $A^i_\pm$ and $J^0$ (dimension 1) and $\psi^a$ (dimension
1/2).
The Kac-Moody algebra defined by the dimension 1 operators is evidently
$SU(2)\times SU(2)\times U(1)$ , which accords with our expectations derived
from the throat geometry.

The superalgebra whose construction we have outlined above is a particular
example of a one-parameter family of N=4 algebras dubbed the $A_\gamma$
algebras. The only problem with it is that the sigma model analysis of
extended supersymmetry (see for example \vnew ) makes quite clear that
the canonically defined supercharges and energy-momentum tensor must satisfy
the standard N=4 algebra, which closes on $T$, $G^a$ and a single level-one
$SU(2)$ Kac-Moody algebra $J^i$. The supercharges defined above obviously do
not have that property. However, if we `improve' them as follows
\eqn\improv{\tilde T=T-{1\over \sqrt{n+2}}\partial J^0 \qquad
             \tilde G^a= G^a-{1\over\sqrt{ n+2}}\partial\psi^a ~~,}
we can show that $\tilde T$ , $\tilde G^a$ and $A^i_-$ (the level-one Kac-Moody
current) close on themselves and enjoy precisely the standard N=4
superalgebra. This says that the full algebra has the standard algebra as a
subalgebra, perhaps no great surprise.

This improvement has a simple physical interpretation: $J^0$ generates a $U(1)$
symmetry which can be regarded as a translation in a free coordinate $\rho$
(that is, we can write $J(z)\sim\p_z\rho(z)$ where $\rho$ is a free scalar
field). The original algebra \gope\ makes no reference to the dilaton and
corresponds physically to a constant dilaton field. It is well-known that,
if one turns on a dilaton which is {\it linear} in a free coordinate $\rho$,
this has the effect of adding a term proportional to $\p^2_z\phi\sim\p_zJ^0(z)$
to $T(z)$ and shifting the central charge of the superconformal algebra by
a constant. With a little care we can show that the linear dilaton implicit
in \improv\ is precisely what we obtained earlier in this section in our
discussion of the WZW-Feigin-Fuks conformal field theory of the throat.
This is a further piece of evidence that the improved energy-momentum tensor
$\tilde T$ is the physically relevant one. Now comes the miracle: $T$ is, in
any event, not physically acceptable because it has a central charge of
$6(n+1)/(n+2)$. The central charge of $\tilde T$ , however, can easily be
shown to be 6, precisely the required value!

This shows that there is an exact conformal field theory of just the right
central charge associated with the throat geometry and verifies the key
r\^ole of N=4 extended supersymmetry in establishing the physics of the model.
There are many fivebrane applications of this exact throat conformal field
theory which are just beginning to be worked out. Perhaps the most
interesting concern the vertex operators of excitations about the wormhole
throat,
among which one must find the moduli of the exact solutions. In any event,
this line of argument shows that the dramatic consequences of (4,4)
superconformal symmetry, which we first extracted from perturbative
considerations, seem to have nonperturbative status.

\newsec{Other Supersymmetric Solitons}
For completeness we will briefly mention in this section some of the other
supersymmetric string solitons which have been discussed in the literature.
These solitons are mainly understood from a spacetime point of view; a
worldsheet analysis of the type discussed in this paper remains to be
performed.

In \refs{\DAHA,\DGHR}, supersymmetric one-brane solutions were
found. The construction of this solution involves an intriguing interplay
between spacetime and worldsheet methods.
These ``solutions'' do not solve the equations of motion
following from \action\ in the usual sense,
rather there is a singularity at the origin corresponding to the
presence of a zero-thickness fundamental string. The form of this
singularity is then precisely dictated by the known coupling of spacetime
vertex operators to the string worldsheet.

In \HASTR\ a one-brane solution of heterotic string theory
was found which is an
everywhere smooth solution of the equations of motion of \action.
The construction of this solution involves crucially
the properties of octonions. One of the many bizarre
features of this soliton is that it breaks 15 of the 16
supersymmetries of ten-dimensional Minkowski space, in
contrast to previously known examples of supersymmetric solitons
which all break half of the supersymmetries. Clearly this is the
odd duck in the family of supersymmetric string solitons.
The fact that this solution does not have finite energy per unit length
further complicates its physical implications.

The fivebrane solutions discussed in these lectures are characterized
by a nonvanishing integral of the three-form field strength $H$,
while the abovementioned string solutions carry a non-zero integral
of the dual of $H$.
In type II strings, there are a variety of $d$-form field strengths.
These fields are not present in the heterotic string because they
are created by Ramond-Ramond vertex operators. For each field
there exists \bhole\ a $p$-brane solution carrying the corresponding
charge or its dual. A conformal field theoretic description of
these solutions is a challenging and interesting problem because
of the non-trivial Ramond-Ramond backgrounds.

The possibility has often been contemplated that our universe
is in fact a three-brane embedded in a higher-dimensional
space. The difficulty with this idea is the opposite of the
difficulty associated with Kaluza-Klein compactification:
there are too few
solutions rather than too many, and none with the right properties
have been found. In particular, although it is easy to obtain
scalar and spin $1/2$ zero modes on such a membrane, it has not been possible
until recently to find theories with higher spin zero modes.
In \wbran\ it was shown that there do exist five-brane solutions of
type II string theory which possess not only spin zero and spin $1/2$
zero modes but also spin one zero modes. Unfortunately, the construction
presented
there also suggests that spin two zero modes cannot be found by the same
mechanism.
As a perhaps more realistic example, the $p$-brane solutions of \bhole\  do in
fact
contain a three-brane solution\dflusix. It is
associated with the self-dual five-form of the chiral
IIb theory. The effective four-dimensional theory has extended
$N=4$ supersymmetry. This may be a useful model for exploring this
alternate method of getting rid of extra dimensions.

Finally there is the issue of solitons in the context of
string compactification to four dimensions\refs{\toptools, \foursol}. This is a
rich subject--too rich to review here. One hope is that
`realistic' string solitons may have peculiar enough
properties that their detection could serve as a
'smoking gun' for string theory. For example in \dine\
it is argued that the fractional charge carried by
certain string solitons could serve this purpose. Of
course direct detection of string solitons is generally difficult
because they are typically ultramassive, but perhaps there
are surprises in store.

\newsec{Conclusion}

In these lectures, we have constructed a special set of conformal field
theories which have the interpretation of soliton solutions of heterotic
string theory. We first constructed them as solutions of the leading
order in $\apm$ beta function conditions and then showed that, owing to
an extended worldsheet supersymmetry, the associated nonlinear sigma model
is an exact conformal field theory. It is the existence of an explicit and
exact conformal field theory associated with the soliton solution which
distinguishes the solution described here from previous attempts to
construct string theory solitons. There are several lines of inquiry which
can be pursued now that ``exact'' string solitons exist. One issue concerns
the mass of the soliton. In all previous discussions of string solitons,
the mass has been computed using the lowest-order spacetime effective action
\action. It would obviously be desirable to have a purely
conformal field theoretic definition of the mass--perhaps in terms of some
correlation functions. Our exact soliton conformal field
theory should provide a useful
context for addressing this question. A second issue is
the question of stringy collective coordinates and their
semiclassical quantization. It should be an instructive challenge to
translate the well-known standard field theory physics of collective
coordinates into the string theory context. This is a nontrivial
matter because motion in collective coordinate space becomes motion in a
space of conformal field theories and it is a nontrivial matter to find
the action associated with such motions (and knowing the exact underlying
conformal field theories should help). Yet another question to pursue is
that of stringy black hole physics. We noted that our solitons
were similar to the extreme Reissner-Nordstrom black holes in the sense that,
while they have no singularity or event horizon, if one increases their mass
by any finite amount (while keeping the axion charge fixed), an event horizon
and a singularity (lying at a finite geodesic distance from any finite point)
will appear. Such black hole solitons can easily be created by scattering
some external particle on an extremal soliton and, by studying stringy
scattering theory about the extremal soliton, one should be able to explore,
in a controlled way, how a stringy black hole Hawking radiates and the nature
of the final state it approaches. These are quite difficult questions, but
having precise control of the underlying conformal field theory may allow
us to make progress on them. Perhaps it will be possible to report on
progress along these lines at the next Trieste School.

\bigbreak \bigskip \bigskip \centerline{{\bf Acknowledgements}} \nobreak
We would like to thank Emil Martinec for collaboration on the material
contained in section 8. This work was supported in part by DOE grants
DE-AC02-84-1553 and 8-48061-2500-3, and by NSF grant PHY90-00386. J.H. also
acknowledges the support of a NSF PYI award PHY91-96117.
Some of the material in these lectures was presented by C.G.C. at the Sixth
J.~A.~Swieca School of Theoretical Physics (January 1991) \swieca.

\listrefs

\bye